\documentclass[iop]{emulateapj}
\shortauthors{Sekanina \& Kracht}
\shorttitle{SOHO/STEREO Kreutz Sungrazers and Comet C/2011 W3}
\slugcomment{Version \today}
\newcommand{\Rsun}{$R_{\mbox{\scriptsize \boldmath $\odot$}}$}

\begin{document}
\title{Population of SOHO/STEREO Kreutz Sungrazers and\\the
Arrival of Comet C/2011 W3 (Lovejoy)}
\author{Zdenek Sekanina$^1$ and Rainer Kracht$^2$}
\affil{$^1$Jet Propulsion Laboratory, California Institute of Technology,
4800 Oak Grove Drive, Pasadena, CA 91109, U.S.A. \\
$^2$Ostlandring 53, D-25335 Elmshorn, Schleswig-Holstein, Germany}
\email{Zdenek.Sekanina@jpl.nasa.gov, r.kracht@t-online.de}

\begin{abstract}
\noindent
We examine properties of the population of SOHO/STEREO (dwarf) Kreutz
sungrazing comets from 2004 to 2013, including the arrival rates, peculiar
gaps, and a potential relationship to the spectacular comet C/2011 W3
(Lovejoy).  Selection effects, influencing the observed distribution,
are largely absent among bright dwarf sungrazers, whose temporal sequence
implies the presence~of~a swarm, with the objects brighter at maximum than
apparent magnitude~3 arriving at a peak~rate~of $\sim$4.6 per year in late
2010, while those brighter than magnitude 2 at a peak rate of $\sim$4.3 per
year in early 2011, both a few times the pre-swarm rate.  The entire
population of SOHO/STEREO Kreutz sungrazers also peaked about one year
before the appearance of C/2011~W3.  Orbital data show, however, that a
great majority of bright dwarf sungrazers moved in paths similar to that of
comet C/1843~D1, deviating 10$^\circ$ or more from the orbit of C/2011~W3
in the angular~elements.  The evidence from the swarm and the overall
elevated arrival rates suggests the existence of a fragmented sizable
sungrazer that shortly preceded C/2011~W3 but was independent of it.  On
the other hand, these findings represent another {\it warning signal\/}
that the expected 21st-century cluster of spectacular Kreutz comets is on
its way to perihelion, to arrive during the coming decades.  It is only in
this sense that we find a parallel link between C/2011~W3 and the spikes
in the population of SOHO/STEREO Kreutz sungrazers.
\end{abstract}

\keywords{comets: general --- comets: individual (X/1106 C1, C/1843 D1,
D/1993 F2, C/1996 Y1, C/2001~G2, C/2003 F5, C/2003 K7, C/2004 A4, C/2004 P5,
C/2005 U5, C/2006 A5, C/2006 V2, C/2006 U8, C/2007 S4, C/2008 K4, C/2009 C3,
C/2009 D4, C/2009 Y4, C/2010 B3, C/2010 C4, C/2010 E6, C/2010 G4, C/2010 U8,
C/2010 V8, C/2010 W2, C/2010~X11--X17, C/2010 Y1--Y16, C/2011 N3, C/2011 W3,
C/2012 E2, C/2013 F4, SOHO-2062, -2072, -2143, -2505, -2571, -2574) ---
methods: data analysis{\vspace{-0.15cm}}}

\section{Introduction}

The osculating orbital period of about 700 years determined for comet
C/2011 W3 (Lovejoy), the most recent spectacular member of the Kreutz system
of sungrazers (Sekanina \& Chodas 2012), rules out this comet's identity with
any known Kreutz sungrazer.  In particular, it cannot represent the return
of any member of this system that has arrived to perihelion since the 17th
century (Kreutz 1901, Marsden 1967), thus providing another example of the
immense complexity of this~comet~\mbox{family}.  With about 2000 minor members
known, thanks to the ongoing vigorous search by amateur astronomers in images
exposed with the coronagraphs on board the {\it \mbox{Solar} and Heliospheric
Observatory\/} (SOHO; see Brueckner et al.\ 1995) and, more recently,
also with the corona\-graphs on board the two spacecraft of the {\it Solar
Terrestrial Relations Observatory\/} (STEREO; see Howard et al.\ 2008),
the relationship between the minor sungrazers and comet C/2011~W3 needs to be
addressed.  For example, did the temporal distribution of the SOHO/STEREO
Kreutz sungrazers change on account of the impending arrival of C/2011 W3?
Is there any evidence for changes in the brightness distribution of these
minor sungrazers?  If yes, how could any such changes be interpreted?  And
could the results be employed in the future to forecast the {\it imminent\/}
arrival of another spectacular sungrazer?

\section{Bright SOHO/STEREO Sungrazers in 2004--2013}

The prediction, published several years ago (Sekanina \& Chodas 2007), of
another cluster of bright sungrazers to appear in the coming decades was
already commented on (Sekanina \& Chodas 2012).  Based on a study of the
long-term evolution of the Kreutz system, this prediction was borne out by
C/2011 W3, the expected cluster's first apparent member.  In comparison,
a successful effort aimed at forecasting the arrival of a strikingly bright
sungrazer we refer to at the end of Sec.~1 is meant to confine the object's
appearance to a temporally much more constrained window.

To address the raised questions, we begin with the distribution, between
2004 and 2013, of arrival times of the SOHO/STEREO Kreutz sungrazers that
were at their maximum brighter than apparent magnitude 3; they are referred
to hereafter as the {\it bright dwarf\/} sungrazers.  Like the fainter Kreutz
minicomets, they all failed to survive perihelion.  We felt that the interval
of nearly eight years is more than sufficient to cover all interesting
objects in the SOHO/STEREO population that preceded, yet might still show a
potential relationship to, C/2011~W3.

We should emphasize that because the definition of a bright dwarf sungrazer is
entirely arbitrary, the limit at apparent magnitude 3 should be regarded only
as a crude constraint.  Indeed, because of SOHO's and STEREO's instrumental
limitations and biases as well as operational practices and options (e.g.,
orange vs.\ clear filters, variable image cadence, uneven duty cycle, etc.;
see Knight et al.\ 2010 for a comprehensive photometric analysis), it could
be very difficult to determine a magnitude with much accuracy and virtually
impossible to establish its equivalent value in any standard photometric band.
For example, to avoid mixing the clear-filter and orange-filter magnitudes
for objects observed with SOHO's C2 and/or C3 coronagraphs is practically
unfeasible, because this problem is intrinsic to the data-acquisition process.
The resulting color index is known to show an enormous scatter of several
magnitudes (Knight et al.\ 2010) and even though systematic trends are vaguely
detected, a color correction inferred from the statistics can never be
warranted to indiscriminately apply to any particular sungrazer's magnitude
and is not necessarily better than no correction.  Among the brightest comets
listed by Knight et al., such as C/1998~K10, C/2000~H2, or C/2003~K7, the
color index is indeed near zero.

The objects that satisfy our definition of a bright dwarf sungrazer are
summarized in Table~1, which presents 19~selected objects.  The magnitude
of most is at maximum brightness in the C2 coronagraph, even though some
may have peaked in C3, with no such data being available.  Only for a few
listed sungrazers the brightness was measured in C3.  The magnitudes were
taken from various data sources, as indicated.  Because of uncertainties,
borderline cases are unavoidable.  It is always possible that one or two
listed objects may, in a uniform photometric system, be somewhat fainter
than the chosen magnitude limit, while the brightness of a very few comets
not included may just barely exceed it.  In any case, we regard the set as
a fairly representative sample.

\begin{table}
\noindent
\begin{center}
{\footnotesize {\bf Table 1}\\[0.08cm]
{\sc SOHO/STEREO Kreutz Sungrazers from 2004--2013\\Brighter at Maximum
 Light Than Magnitude 3.}\\[0.08cm]
\begin{tabular}{l@{\hspace{0.25cm}}l@{\hspace{0.05cm}}c@{\hspace{0.1cm}}l}
\hline\hline\\[-0.27cm]
       & Date of first & Apparent  & \\[-0.05cm]
       & observation   & magnitude & \\[-0.04cm]
Object & in C3 (UT) & at max. & Reference(s)$^{\rm a}$ \\[0.03cm]
\hline\\[-0.22cm]
C/2006 A5 & 2006 Jan.\,4  & 2.9\rlap{$^{\rm b}$} & IAUC 8694,\,[1] \\
C/2006 V2 & 2006 Nov.\,1  & 1.3        & IAUC 8811,\,[1] \\
C/2008 K4 & 2008 May 22   & 0.5        & IAUC 8982,\,[1] \\
C/2009 C3 & 2009 Feb.\,4  & 2--3       & IAUC 9055 \\
C/2009 D4 & 2009 Feb.\,21 & 2--3       & IAUC 9056 \\
C/2009 Y4 & 2009 Dec.\,30 & 1.0        & IAUC 9117,\,[1],\,[2] \\
C/2010 B3 & 2010 Jan.\,17 & $\sim$2$^{\rm c}$ & [2] \\
C/2010 E6 & 2010 Mar.\,10 & 1.1        & IAUC 9151,\,9157,\,[1],\,[2] \\
C/2010 G4 & 2010 Apr.\,8  & 0          & [1],\,[2] \\
C/2010 U8 & 2010 Oct.\,19 & 1.8        & [1],\,[2] \\
C/2010 V8 & 2010 Nov.\,12 & 2.3        & [1],\,[2] \\
C/2010 W2 & 2010 Nov.\,17 & 1.4        & [1],\,[2] \\
SOHO-2062 & 2011 May 9    & 0.9        & [1],\,[2] \\
SOHO-2072 & 2011 May 20   & 1.5        & [1] \\
C/2011 N3 & 2011 Jul.\,4  & 1          & IAUC 9227,\,[1],\,[2] \\
SOHO-2143 & 2011 Sept.\,30 & $-$0.5    & [1],\,[2],\,[3] \\
C/2012 E2 & 2012 Mar.\,13 & 1          & CBET 3047,\,[1],\,[3] \\
SOHO-2505 & 2013 May 10   & 1.8        & [1] \\
SOHO-2571 & 2013 Aug.\,18 & 2.6        & [1] \\[0.03cm]
\hline\\[-0.24cm]
%
%
\multicolumn{4}{l}{\parbox{8.2cm}{$^{\rm a}$\,\scriptsize Website{\vspace{-0.04cm}} URLs:\ [1]\,{\tt http:/{\hspace{-0.04cm}}/www.rkracht.de/soho/bright/bright.htm}; [2] {\tt http:/{\hspace{-0.05cm}}/remanzacco.blogspot.com} for 2010 dates
of Jan.\,2, {\vspace{-0.04cm}}Jan.\,23, Mar.\,12, Apr.\,10, Oct.\,21, Nov.\,16,
and Nov.\,20, and for{\vspace{-0.04cm}} 2011 dates of May 11,  Jul.\,5, and
{\vspace{-0.04cm}}Oct.\,2; [3] {\tt
http:/{\hspace{-0.05cm}}/sungrazer.nrl.navy.mil}, news items dated Oct.\,4,
2011 and Mar.\,16, 2012.}} \\[0.45cm]
\multicolumn{4}{l}{\parbox{8.2cm}{$^{\rm b}$\,\scriptsize Claim on IAUC 8694
of this comet's {\vspace{-0.04cm}}peak magnitude of ``perhaps $-$1.5'' is
strongly disputed in [1].}}\\[0.14cm]
\multicolumn{4}{l}{\parbox{8.2cm}{$^{\rm c}$\,\scriptsize This crude estimate
is based {\vspace{-0.04cm}}on a statement in [2] that this comet was
$\sim$1--1.5 magnitudes fainter {\vspace{-0.04cm}}than C/2009 Y4.  Comparison
of C3 images shows that C/2010 B3 was clearly {\vspace{-0.04cm}}brighter
than C/2007 C6, which according to IAUC 8844 peaked at magnitude
$\sim$3.}}\\[-0.1cm]
%
%
\end{tabular}}
\end{center}
\end{table}

We took further measures to constrain the magnitudes for the individual entries
in Table 1.  Some of the objects were compared with other sungrazers, whose
maximum brightness was reported to have been near apparent magnitude~3, by
inspecting the SOHO's C2 and C3 coronagraphic movies.\footnote{See {\tt
http:/{\hspace{-0.05cm}}/sohodata.nascom.nasa.gov/cgi-bin/data\_query.}}
Known magnitudes for these comparisons were taken primarily from Knight
et al.\ (2010).  The second author's website blog,\footnote{See
{\tt http:/{\hspace{-0.05cm}}/www.rkracht.de/soho/bright/bright.htm.}} which
provides a table of bright Kreutz sungrazers, proved especially useful in
compiling Table~1.  Also, his determination that the detector of the SOHO's
C2 coronagraph begins to saturate by objects of apparent magnitude $\sim$1.5
allows one to instantly set an upper or lower brightness limit by just looking
at the image.  The arguments for including some objects are presented in the
table's footnotes, which also provide information on the electronic blogs.
Finally, the properties of the chosen set of SOHO/STEREO sungrazers brighter
at maximum light than apparent magnitude~3 is in the following compared
with the properties of a subset, the sungrazers brighter at maximum light
than magnitude~2.  The magnitudes of most of these objects were determined
from C2 imaging and come from the second author's website.$^2$

While the number of bright SOHO/STEREO sungrazers is limited, their advantage
is that they are practically free from the selection effects that influence
the detection of fainter Kreutz minicomets.  In particular, strong annual
double-peak variations in the discovery rate from images taken with the C2
coronagraph are due to periodic orientation changes in the Kreutz comets'
orbital plane relative to Earth (Sekanina 2002; Knight et al.\ 2010):\ the
more prominent peak occurs in May, while the less pronounced but much more
extended one covers October-December (e.g., Knight et al.'s Figure 13).
One of the deep minima is in January and, accordingly, in a set of 19 objects
there should be none during that month.  Yet, in Table~1 we have two of them,
consistent with an essentially random annual-distribution sample.

\begin{figure}[t]
\vspace{-9.3cm}
\hspace{1.37cm}
\centerline{
\scalebox{0.74}{
\includegraphics{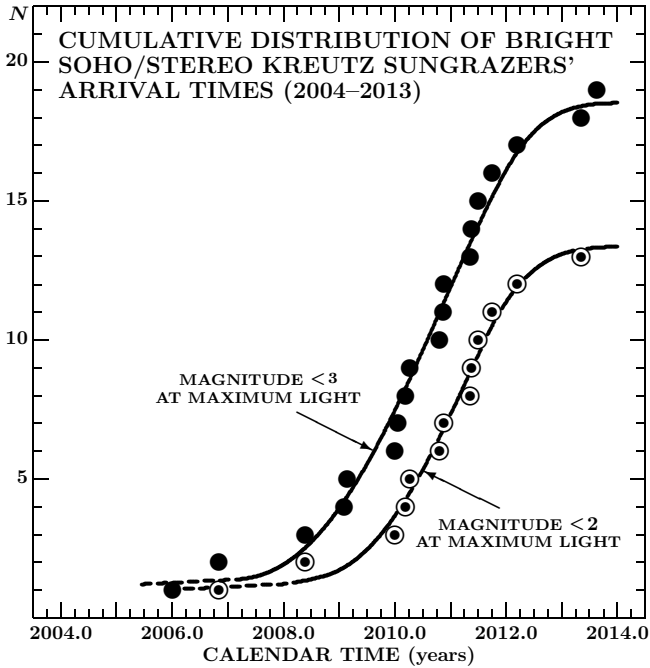}}} 
\vspace{-4cm}
\caption{Cumulative distributions of the bright dwarf Kreutz sungrazers
observed with the SOHO and/or STEREO coronagraphs in 2004--2013.  The solid
circles refer to 19 objects brighter at their maximum than magnitude 3, the
circled dots to 13 objects brighter than magnitude 2.  The data points are
from Table~1.  The curves are the fitted 4th-power polynomials slightly
modified at the ends, where the distribution curves are poorly
defined.{\vspace{0.2cm}}}
\end{figure}

For the years 2004--2013, the cumulative distributions of arrival times of the
SOHO/STEREO sungrazers brighter at their maximum than, respectively, apparent
magnitude~3 and 2 are plotted in Figure~1, in which the entries from Table~1
are fitted with a polynomial of the 4th power modified at the ends.  The
arrival time is the calendar date of the first observation in SOHO's C3
coronagraph, which is shown in column~2 of Table~1.  Because no Kreutz
sungrazer can enter the field of view of the SOHO's C3 coronagraph earlier
than less than 5 days --- and the field of view of the C2 coronagraph earlier
than less than 0.5 day --- before perihelion, the difference between the
arrival time and the perihelion time is negligible on the scale of Figure~1.
The reason for choosing this approximation is the fact that perihelion times
of nearly all sungrazers that have arrived since the beginning of 2011 are
not yet available.

The upper curve in Figure 1 illustrates the rapidly increasing number of
bright SOHO/STEREO sungrazers starting in 2008, while the lower curve, based
on 13 objects, a year or two later.  The overall shapes of the two distribution
curves are similar, except for the slightly less steep rate of growth of the
lower curve, as expected.  This fact strengthens our confidence that we deal
with representative sets of bright dwarf sungrazers.  Both distributions ---
especially the upper one --- do, however, display pronounced local deviations
from the fitted polynomials, appearing as step-like features separated from
each other by gaps extending over several months.  Their nature will be
addressed in Sec.~5.

No sungrazers were brighter at maximum light than magnitude~3 in 2004--2005,
although in 2003, just off the scale of Figure~1, two were brighter:\
C/2003~F5 and C/2003~K7.  From Table 2 of Knight et al.\ (2010), which
includes only those Kreutz sungrazers whose brightness peaked in C3, it
follows that during the decade 1996--2005 there was a total of eight SOHO
sungrazers brighter at maximum than magnitude 3, suggesting an average
rate of 0.8 per year.  Knight et al.'s list does not though include
C/1996~Y1 and C/2001~G2, both of which according to Kracht (see footnote
2) were brighter at maximum than magnitude~1.  For the same period of time,
Kracht lists eight sungrazers --- or an average of 0.8 per year --- that,
mostly in C2, reached at maximum at least apparent magnitude~1, the same
rate as that implied by Knight et al.'s data but for clearly brighter
objects.

\begin{figure}
\vspace{-11.93cm}
\hspace{1.01cm}
\centerline{
\scalebox{0.815}{
\includegraphics{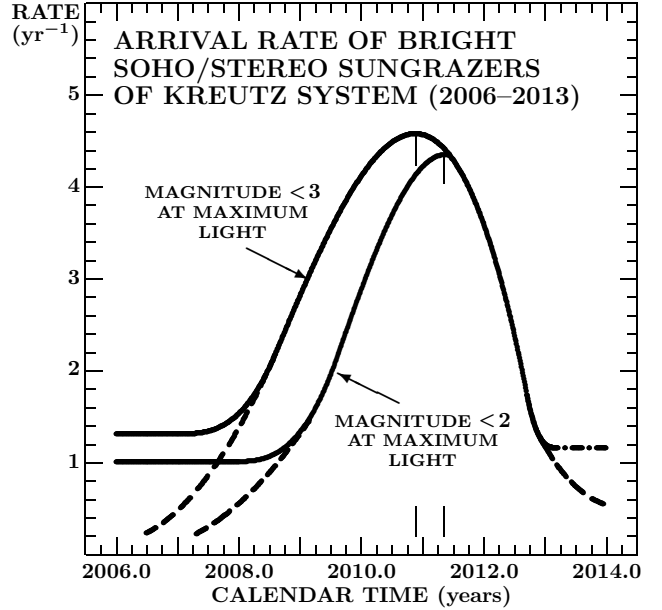}}} 
\vspace{-4.45cm}
\caption{Variations in the arrival rate of the bright dwarf Kreutz
sungrazers observed with the SOHO and STEREO coronagraphs in 2004--2013.
The curves are the differentiated polynomials fitting the cumulative
distributions in Figure 1.  The peak arrival rates are, respectively,
$\sim$4.6 per year at 2010.88 for the objects brighter at their maximum
than apparent magnitude~3 and $\sim$4.3 per year at 2011.35 for those
brighter than magnitude~2.  The poorly defined, dashed sections of the
curves before 2009 and from late 2012 on are replaced with the estimated
average rates from 1996--2005.  The dot-dashed curve is a ball-park
extrapolation into the future.{\vspace{0.3cm}}}
\end{figure}

By differentiating the polynomial approximations to the curves in Figure 1,
we derived the smoothed arrival-rate distributions of the bright SOHO/STEREO
Kreutz sungrazers in Figure 2.  The arrival-rate curves --- poorly determined
before and in early 2008 and again from late 2012 on --- tend to converge to
zero in 2006 and 2014.

From Figure 2, the times of peak arrival rate are 2010.88 ($\sim$\,mid-November
2010) for the dwarf sungrazers brighter at maximum light than magnitude 3 and
2011.35 ($\sim$\,early May 2011) for those brighter at maximum light than
magnitude 2.  With the larger set assigned a greater weight and with an
estimated uncertainty of two to four months in the times of peak arrival
rate, we conclude that, on the average, this group of {\it bright dwarf
sungrazers preceded the arrival of C/2011 W3 by just about one year, which
suggests a potentially close relationship between both\/}.

The polynomial fitting indicates that the smoothed peak arrival rates were
$\sim$4.6 and $\sim$4.3 objects per year, respectively, for the two sets
of data, showing that the arrival rate of sungrazers with a magnitude at
their maximum between 2 and 3 was only about 0.3 per year.  At much lower
arrival rates before 2009, the difference could not possibly have been any
greater than this.  For illustration, Figure 2 shows the pre-2008 constant
rates of 1.3 per year for the sungrazers brighter at maximum light than
magnitude~3 and 1.0 per year for those brighter than magnitude 2.

The relatively flat magnitude distribution function of the bright SOHO/STEREO
Kreutz sungrazers is also documented by the surprising merger of the two
curves in Figure 2 at times starting in the second half of 2011, suggesting a
deficit of the sungrazers with magnitudes at maximum light between 2 and 3.
Only to a degree may this property of the distribution function be affected
by insufficient data.  The same probably also applies to the precipitous
dropoff rate during 2012, which in Figure 2 appears steeper than the rate
of rise before 2011 and which is of course a product of a deficit of bright
dwarf sungrazers during much of 2012.

\begin{table*}[ht]
\noindent
\begin{center}
{\footnotesize {\bf Table 2}\\[0.08cm]
{\sc Orbital Elements and Apsidal Orientation of Bright SOHO/STEREO Kreutz
Sungrazers from the Years 2004--2013\\Compared with the Orbital
Data for Spectacular Comets C/1843 D1 and C/2011 W3 (Eq.\
J2000.0).}\\[0.08cm]
\begin{tabular}{l@{\hspace{0.35cm}}l@{\hspace{0.07cm}}c@{\hspace{0.07cm}}r@{\hspace{0.55cm}}c@{\hspace{0.45cm}}c@{\hspace{0.4cm}}c@{\hspace{0.4cm}}c@{\hspace{0.5cm}}c@{\hspace{0.45cm}}c@{\hspace{0.3cm}}c@{\hspace{0.15cm}}l}
\hline\hline\\[-0.23cm]
 & \multicolumn{7}{@{\hspace{-0.35cm}}c}{Orbital elements}
 & \multicolumn{2}{@{\hspace{-0.35cm}}c}{Line of apsides} & Number
 & \\[-0.07cm]
 & \multicolumn{7}{@{\hspace{-0.35cm}}c}{\rule[0.6ex]{7.2cm}{0.4pt}}
 & \multicolumn{2}{@{\hspace{-0.35cm}}c}{\rule[0.6ex]{2.5cm}{0.4pt}}
 & of used \\[-0.08cm]
Object$^{\rm a}$
 & \multicolumn{3}{@{\hspace{-0.1cm}}c}{$t_\pi$ (ET)}
 & $\omega$ & $\Omega$ & $i$ & $q$\,($R_{\mbox{\boldmath $\scriptstyle
   \odot$}}$) & $L_\pi$ & $B_\pi$ & positions & Reference \\[0.05cm]
\hline\\[-0.2cm]
C/2006 A5{\boldmath $^\ast$} & 2006 & Jan. & 5.66 & 80$^\circ\!\!$.27
          & 0$^\circ\!\!$.54 & 144$^\circ\!\!$.04 & 0.92 & 282$^\circ\!\!$.50
          & +35$^\circ\!\!$.36 & 62 & MPC\,56609 \\
C/2006 V2 & 2006 & Nov. & 3.78 & 84.76 & 5.89 & 144.58 & 1.07 & 282.31
          & +35.25 & 48 & MPC\,58779 \\
C/2008 K4{\boldmath $^\ast$} & 2008 & May & 23.84\rlap{3} & 82.63\rlap{1}
          & 3.57\rlap{5} & 144.57\rlap{1} & 0.99\rlap{0} & 282.59\rlap{4}
          & +35.09\rlap{3} & \llap{1}39 & MPC\,63377 \\
C/2009 C3{\boldmath $^\ast$} & 2009 & Feb. & 5.65 & 82.15 & 3.59 &144.82
          & 1.01 & 283.16 & +34.80 & 17 & MPC\,66465 \\
C/2009 D4{\boldmath $^\ast$} & 2009 & Feb. & 23.43\rlap{8} & 82.30\rlap{8}
          & 4.20\rlap{9} & 144.67\rlap{7} & 1.19\rlap{2} & 283.60\rlap{8}
          & +34.95\rlap{8} & 99 &  MPC\,66465 \\
C/2009 Y4{\boldmath $^\ast$} & 2010 & Jan. & 3.53\rlap{7} & 83.19\rlap{1}
          & 4.56\rlap{7} & 144.57\rlap{3} & 1.04\rlap{4} & 282.90\rlap{3}
          & +35.14\rlap{0} & \llap{1}28 & MPC\,68393 \\
C/2010 B3{\boldmath $^\ast$} & 2010 & Jan. & 21.78\rlap{3} & 83.17\rlap{2}
          & 4.42\rlap{0} & 144.57\rlap{2} & 1.03\rlap{1} & 282.78\rlap{0}
          & +35.13\rlap{9} & \llap{1}98 & MPC\,72130 \\
C/2010 E6{\boldmath $^\ast$} & 2010 & Mar. & 12.89\rlap{3} & 83.20\rlap{6}
          & 4.38\rlap{1} & 144.59\rlap{5} & 1.03\rlap{1} & 282.69\rlap{7}
          & +35.12\rlap{0} & \llap{1}41 & MPC\,70816 \\
C/2010 G4{\boldmath $^\ast$} & 2010 & Apr. & 10.24\rlap{9} & 83.40\rlap{9}
          & 4.39\rlap{0} & 144.66\rlap{8} & 1.01\rlap{8} & 282.45\rlap{1}
          & +35.06\rlap{4} & \llap{1}07 & MPC\,72133 \\
C/2010 U8{\boldmath $^\ast$} & 2010 & Oct. & 21.48\rlap{1} & 85.08\rlap{7}
          & 5.48\rlap{0} & 144.64\rlap{1} & 1.07\rlap{9} & 281.49\rlap{7}
          & +35.21\rlap{0} & 83 & This work \\
C/2010 V8{\boldmath $^\ast$} & 2010 & Nov. & 14.50\rlap{7} & 85.18\rlap{1}
          & 6.57\rlap{1} & 144.49\rlap{2} & 1.15\rlap{9} & 282.48\rlap{4}
          & +35.36\rlap{4} & 61 & This work \\
C/2010 W2{\boldmath $^\ast$} & 2010 & Nov. & 19.24\rlap{5} & 85.24\rlap{7}
          & 6.66\rlap{3} & 144.45\rlap{8} & 1.15\rlap{7} & 282.49\rlap{7}
          & +35.40\rlap{1} & 77 & This work \\
SOHO-2062 & 2011 & May  & 11.26\rlap{9} & 84.72\rlap{6} & 6.75\rlap{6}
          & 144.11\rlap{3} & 1.35\rlap{2} & 283.25\rlap{6} & +35.71\rlap{7}
          & \llap{1}57 & This work \\
SOHO-2072{\boldmath $^\ast$} & 2011 & May  & 22.05\rlap{4} & 85.33\rlap{5}
          & 6.83\rlap{6} & 144.40\rlap{5} & 1.16\rlap{0} & 282.56\rlap{6}
          & +35.45\rlap{9} & \llap{1}35 & This work \\
C/2011 N3{\boldmath $^\ast$} & 2011 & Jul. & 6.00 & 85.10 & 6.41 & 144.41
          & 1.14 & 282.43 & +35.44 & \llap{1}02 & MPC\,75517\\
SOHO-2143{\boldmath $^\ast$} & 2011 & Oct. & 1.85\rlap{3} & 85.26\rlap{7}
          & 6.62\rlap{3} & 144.47\rlap{0} & 1.14\rlap{0} & 282.43\rlap{2}
          & +35.39\rlap{1} & \llap{1}29 & This work \\
C/2012 E2{\boldmath $^\ast$} & 2012 & Mar. & 15.00\rlap{6} & 85.72\rlap{8}
          & 7.46\rlap{1} & 144.34\rlap{4} & 1.15\rlap{3} & 282.71\rlap{4}
          & +35.54\rlap{2} & 61 & This work \\
SOHO-2505{\boldmath $^\ast$} & 2013 & May  & 12.31\rlap{0} & 85.78\rlap{2}
          & 7.13\rlap{3} & 144.62\rlap{1} & 1.10\rlap{9} & 282.30\rlap{2}
          & +35.26\rlap{9} & \llap{1}28 & This work \\
SOHO-2571 & 2013 & Aug. & 20.11\rlap{8} & 76.70\rlap{8} & \llap{35}6.27\rlap{7}
 & 143.69\rlap{5} & 1.29\rlap{0} & 282.50\rlap{7} & +35.20\rlap{0}
 & \llap{1}75 & This work \\[0.02cm]
\hline \\[-0.25cm]
C/1843 D1 & 1843 & Feb. & 27.91\rlap{4} & 82.75\rlap{6} & 3.69\rlap{5}
 & 144.38\rlap{4} & 1.17\rlap{3} & 282.58\rlap{2} & +35.28\rlap{9} & 37
 & Sekanina \& Chodas (2008) \\
C/2011 W3 & 2011 & Dec. & 16.01\rlap{2} & 53.51\rlap{0} & \llap{3}26.369
 & 134.35\rlap{6} & 1.19\rlap{3} & 282.98\rlap{4} & +35.08\rlap{8}
 & \llap{1}23 & Sekanina \& Chodas (2012) \\[0.02cm]
\hline \\[-0.28cm]
\multicolumn{12}{l}{\parbox{15cm}{$^{\rm a}$\,\scriptsize Members of the
 swarm that are particularly tightly related to one another are marked
 with an asterisk.}}\\[0.05cm]
\end{tabular}}
\end{center}
\end{table*}

A major conclusion from our analysis of the temporal distribution of bright
dwarf sungrazers is that in the past several years we witnessed the {\it
arrival of a swarm\/}, or, because of the local deviations from a smooth
curve, the {\it arrival of a swarm of clumps\/} of these objects.  An
exciting possibility exists that all, or at least most, of these bright
minicomets were closely related to each other, and possibly to C/2011~W3.
First, however, this hypothesis needs to be tested dynamically.

\section{Orbital Data for Bright Dwarf Sungrazers}

Parabolic orbital elements for the first nine sungrazers in Table~1 were
computed by B.\ G.\ Marsden and those for C/2011~N3 by G.\ V.\ Williams;
they were published in various {\it Minor Planet Circulars\/} (MPC).  A
parabolic orbit determination for C/2012~E2 in MPC\,79023 rested on merely
39~observations of low accuracy from images taken with the HI1-B and C3
coronagraphs on board, respectively, STEREO-B and SOHO.  With no orbits
available for the more recent objects in Table~1 at the time we began
to tackle this problem, the second author took up the task of measuring
hundreds of their positions and computing their parabolic orbits, including
the orbital re-determination of C/2012~E2 based, in addition, on more
accurate positions from 19~images taken with the COR2-B coronagraph.  Only
after this work was completed did Gray (2013) independently publish the
elements for comets C/2010~U8 (SOHO-1932), C/2010~V8 (SOHO-1948), and
C/2010~W2 (SOHO-1954).

The parabolic orbits for the 19 sungrazers are, with the references, presented
in Table~2; the second author's results are listed for nine.  For comparison,
the last two lines contain the elements from elliptical solutions for two
major sungrazers.  The columns $L_\pi$ and $B_\pi$ provide, respectively, the
longitude and latitude of perihelion.

The orbital data in Table 2 have major implications.  They primarily show an
utter lack of orbital similarity between the 19 bright dwarf sungrazers
on the one hand and C/2011~W3 on the other.  The discrepancies are
$\sim$10$^\circ$ in the inclination and much more in the other angular
elements.  Thus, {\it the potential correlation suggested by the
near-coincidence between the appearance of C/2011~W3 and the arrival-rate
peak of the 19~bright dwarf sungrazers is strongly contradicted by this
orbital evidence\/}.

Our second finding is that except for the last entry, SOHO-2571, the bright
dwarf sungrazers move in orbits that are {\it remarkably similar to that of
C/1843~D1\/}, one~of the two most spectacular Kreutz system's members since
1800, just as do the orbits of most SOHO Kreutz sungrazers (sometimes called
{\it Subgroup I\/}, e.g., Sekanina 2002).

And third, with the exception of two more sungrazers, C/2006~V2 and
SOHO-2062, three elements --- the argument of perihelion $\omega$, the
longitude of the ascending node $\Omega$, and the perihelion distance $q$
--- of the remaining objects' orbits show a {\it strong tendency to increase
systematically with time\/}, despite noise in the data.  These {\it
16~bright dwarf sungrazers\/}, marked with asterisks in Table~2, make
up a {\it swarm of objects that are particularly tightly related to
one another\/} (a tightly-knit swarm).  The ranges spanned are about
5$^\circ\!$.5 in $\omega$, 7$^\circ$ in $\Omega$, and 0.24\,{\Rsun} in
$q$.  The range that measures a systematic trend in the inclination $i$
must be less than 0$^\circ\!$.2.  In fact, the inclination and the
apsidal line exhibit no obvious trends.

\begin{table}[t]
\noindent
\vspace{-0.2cm}
\begin{center}
{\footnotesize {\bf Table 3}\\[0.08cm]
{\sc Average Slopes and Correlation Coefficients for\\Orbital Parameters
of the 16 Tightly Related Members\\of a Swarm of Bright Dwarf Sungrazers
in Table 2.}\\[0.12cm]
\begin{tabular}{l@{\hspace{0.45cm}}c@{\hspace{0.05cm}}c}
\hline\hline\\[-0.25cm]
Orbital & Average & Correlation \\[-0.06cm]
parameter & slope & coefficient \\[0.05cm]
\hline\\[-0.22cm]
Argument of perihelion (yr$^{-1}$)
         & +0$^\circ\!$.889\,$\pm$\,0$^\circ\!$.089 & 0.94 \\
Long.\ ascending node (yr$^{-1}$)
         & +1$^\circ\!$.010\,$\pm$\,0$^\circ\!$.090 & 0.95 \\
Perihelion distance ({\Rsun}\,yr$^{-1}$)
         & +0.0322\,$\pm$\,0.0088 & 0.70 \\
Inclination (yr$^{-1}$) & +0$^\circ\!$.024\,$\pm$\,0$^\circ\!$.027 & 0.23 \\
Longitude of perihelion (yr$^{-1}$)
         & $-$0$^\circ\!$.078\,$\pm$\,0$^\circ\!$.066 & \llap{$-$}0.30 \\
Latitude of perihelion (yr$^{-1}$)
         & +0$^\circ\!$.051\,$\pm$\,0$^\circ\!$.029 & 0.42 \\[0.05cm]
\hline \\[-0.3cm]
\end{tabular}}
\end{center}
\end{table}

To confirm the veracity of these patterns, we offer two tests.  The first
is a comparison of the total range with the estimated uncertainty in each
orbital element.  No errors were published for the orbits in Table~2 taken
from the MPCs.  The solutions for the nine entries computed by the second
author yielded, on the average, the RMS deviations of $\pm$0.0015 day in
$t_\pi$, $\pm$0$^\circ\!$.19 in $\omega$, $\pm$0$^\circ\!$.30 in $\Omega$,
$\pm$0$^\circ\!$.05 in $i$, and $\pm$0.011\,{\Rsun} in $q$.  Equating the
signal with the range and the noise with the sum of the RMS deviations for
the 16 swarm's members calculated as $\sqrt{16} = 4$ times the average RMS
deviation, we have a signal-to-noise ratio of 5--7 for $\omega$, $\Omega$,
and $q$.  On the other hand, the signal-to-noise ratio for $i$ comes out to
be $<$1.  This explains why the systematic trends are apparent in the first
three elements but not in the inclination.

The differences between the behavior of $\omega$, $\Omega$, and $q$ of the
16 swarm members, on the one hand, and their $i$, $L_\pi$, and $B_\pi$, on
the other hand, are independently illustrated by comparing the relative
errors of their average slopes, $\langle d\omega/dt \rangle$, \ldots,
$\langle d B_\pi/dt \rangle$, and the correlation coefficients, which are
summarized in Table 3.  The results show huge gaps in both criteria
between the two groups of orbital parameters.  We will address the meaning
of these effects and provide their interpretation in Sec.~5.

The trends in Table 2 are also corroborated by the object SOHO-2574,
a sungrazer discovered in the second half of August 2013.  From the
brightness measurements by the second author, this comet never attained
apparent magnitude 3, peaking at about 3.8 in C3, so it is not included
in Tables 1 and 2.  However, its orbit, in Table~4, further extends the
time interval of the systematic pattern of the bright dwarf sungrazers.
This fainter object probably is an outlying member of the swarm of tightly
related sungrazers.

\begin{figure*}
\vspace{-2.6cm}
\hspace{-1.5cm}
\centerline{
\scalebox{0.785}{ 
\includegraphics{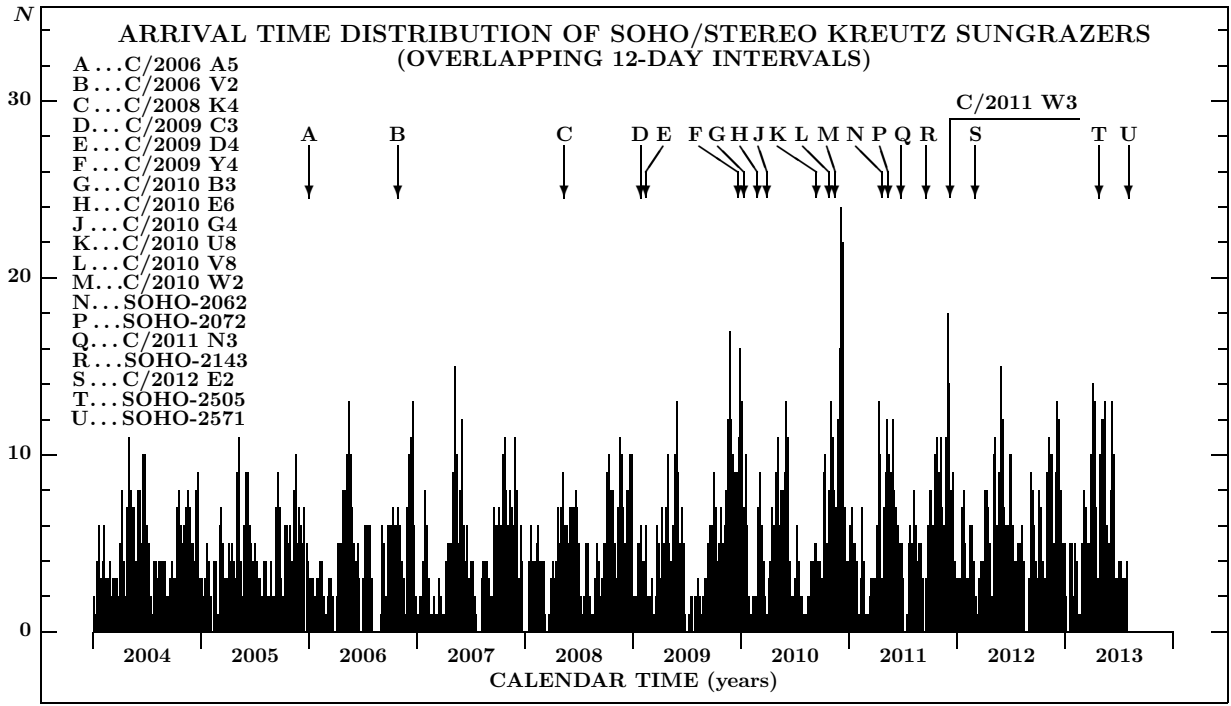}}} 
\vspace{-11.28cm}
\caption{Distribution of arrival times of the Kreutz sungrazers, detected
with the coronagraphs on board the SOHO and/or STEREO spacecraft in the
years 2004--2013.  The sample is arranged into 12-day consecutive intervals,
overlapping each other by six days.  The perihelion time of C/2011~W3 and
the arrival times of the 19 bright dwarf sungrazers (A--U) from Table~1 are
depicted near the top.{\vspace{0.4cm}}}
\end{figure*}

\section{Temporal Distribution of Arrivals of All SOHO/STEREO Sungrazers
in 2004--2013}

\begin{table}[b]
\noindent
\hspace{-0.1cm}
\vspace{0.15cm}
\begin{center}
{\footnotesize {\bf Table 4}\\[0.08cm]
{\sc Parabolic Orbit of Comet SOHO-2574 (Eq.\ J2000.0).}\\[0.04cm]
\begin{tabular}{l@{\hspace{0.1cm}}l@{\hspace{0.45cm}}c@{\hspace{0.15cm}}l@{\hspace{0.45cm}}l@{\hspace{0.1cm}}r}
\hline\hline\\[-0.2cm]
$t_\pi$ & 2013\,Aug.\,23.7201\,ET & $\omega$
        & $\;\:$85$^\circ\!$.817\,$\pm$\,0$^\circ\!$.08 & $L_\pi$
        & 282$^\circ\!$.483 \\
        & \hspace{1.235cm}$\pm$0.0005 & $\Omega$
        & $\;\:\;\:$7$^\circ\!$.356\,$\pm$\,0$^\circ\!$.07 & $B_\pi$ 
        & +35$^\circ\!$.289 \\
$q$     & 1.123\,$\pm$\,0.003\,{\Rsun} & $i$
    & 144$^\circ\!$.603\,$\pm$\,0$^\circ\!$.017 & & 114 obs. \\[0.06cm]
\hline\\[-0.63cm]
\end{tabular}}
\end{center}
\end{table}

The spike-like distribution of arrival times of all Kreutz sungrazers
detected with the coronagraphs on board the SOHO and STEREO spacecraft
from the beginning of 2004 until the time of this paper's submittal in
August 2013 is shown in Figure 3.  The plotted numbers are the totals of
sungrazers with their arrival times within 12-day periods of time with a
six-day overlap.  The first interval (in UT) starts actually on December
29.0, 2003 and ends on January 10.0, 2004, the second interval extends
from January 4.0 until January 16.0, 2004, etc.  The overlap is intended
to suppress sharp variations in the number of arriving sungrazers and
makes thus a search for unexpected peaks deliberately more difficult.

\begin{figure*}
\vspace{-2.7cm}
\hspace{-1.4cm}
\centerline{
\scalebox{0.785}{
\includegraphics{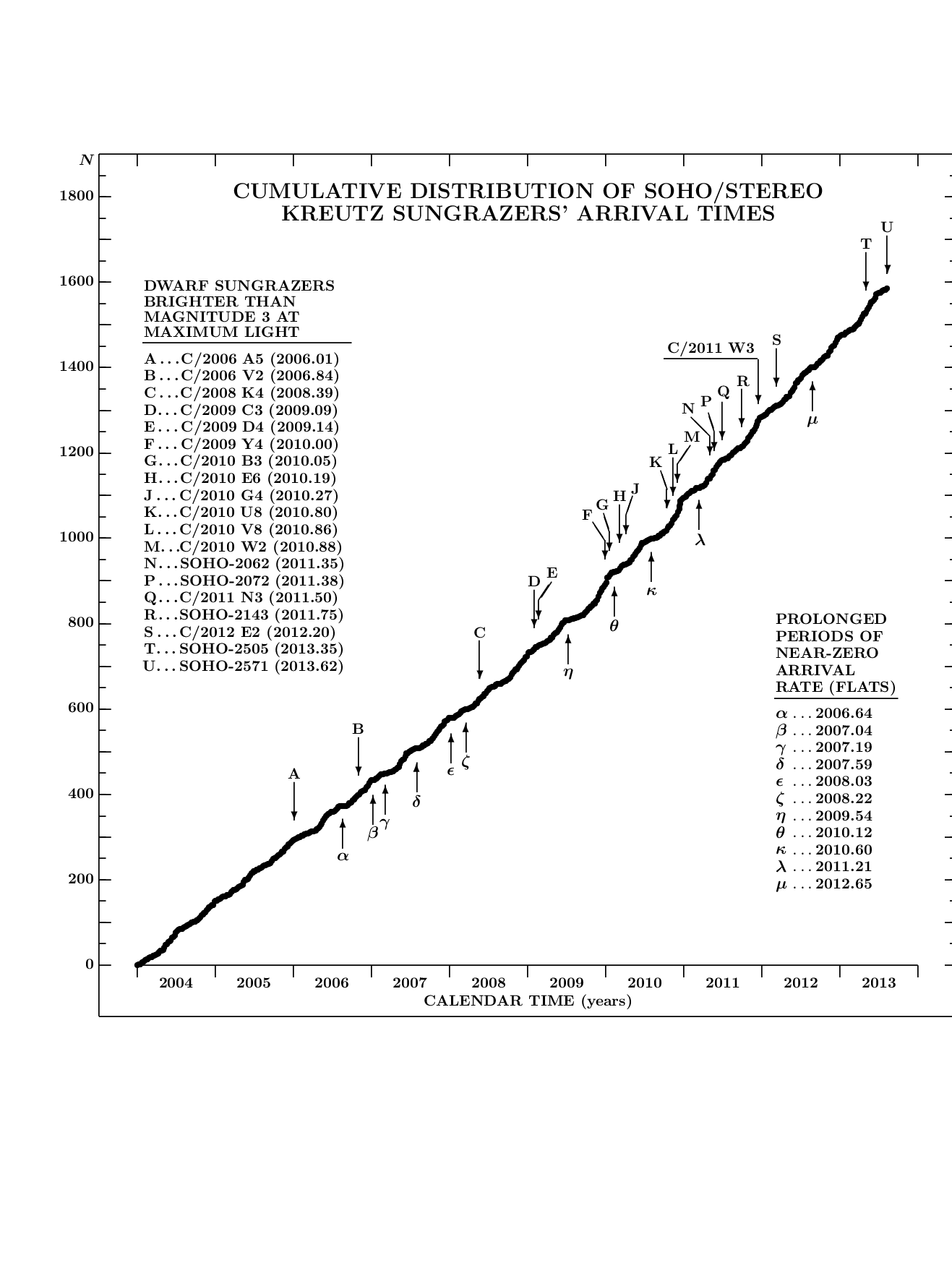}}} 
\vspace{-4.6cm}
\caption{Cumulative distribution of arrival times of the Kreutz sungrazers,
detected with the coronagraphs on board the SOHO and/or STEREO spacecraft
in the years 2004--2013.  The arrival times of the 19 bright SOHO/STEREO
sungrazers from Table 1 are marked near the top of the figure by the
letters A--U.  The perihelion time of C/2011 W3 is also depicted.  The
locations of the 11 flat segments on the curve --- fairly prolonged intervals
of a zero or near-zero arrival rate --- are shown by the Greek letters
$\alpha$--$\mu$.{\vspace*{0.35cm}}}
\end{figure*}

The distribution clearly shows the annual variations due to the selection
effects already noted in Sec.\ 2.  However, superposed on these quasi-periodic
fluctuations are, in spite of the employed overlapping technique, a clear
systematic long-term variation, with the most conspicuous spike at 2010.97,
almost exactly one year prior to the perihelion time of C/2011~W3.  Inspection
suggests that this spike was dominated by {\it a clump of 22 Kreutz
sungrazers\/} detected during eight consecutive days, December 13--20, with an
equivalent rate of an astonishing 1000 objects per year!  The second highest
spike in Figure~3 occurred one year later, at 2011.95, practically coinciding
with the perihelion time of C/2011 W3, and dominated by a clump of 16 sungrazers
discovered over 12 days, December 9--20.  This rate is equivalent to nearly
500 objects per year.  A third, double spike is located symmetrically to the
largest one, in late 2009.  Thus, comparing only December spikes in Figure~3
--- in order to eliminate much of the annual selection effect --- we are
finding a {\it pattern that is strongly resembling that exhibited by the
brighter sungrazers\/} in Figure 2.

\begin{table}[b] 
\noindent
\hspace{-0.1cm}
\begin{center}
{\footnotesize {\bf Table 5}\\[0.1cm]
{\sc Orbital Similarity for 22 Members of December 2010\\Clump of
 SOHO/STEREO Kreutz Sungrazers with\\2010/2011 Swarm of Bright Dwarf
 Comets.}\\[0.1cm]
\begin{tabular}{l@{\hspace{0cm}}c@{\hspace{0.1cm}}c@{\hspace{0.17cm}}c@{\hspace{0.13cm}}c@{\hspace{0cm}}c@{\hspace{0.05cm}}c}
\hline\hline\\[-0.25cm]
 & & & \multicolumn{3}{@{\hspace{0.11cm}}c}{Mean residual from} & \\[-0.05cm]
 & Number & Corona-
 & \multicolumn{3}{@{\hspace{0.11cm}}c}{\rule[0.6ex]{2.9cm}{0.4pt}} & \\[-0.05cm]
 & of used & graph(s) & Best & C/1843 & C/2011 & Part of \\[-0.05cm]
Object$^{\rm a}$ & positions & used & fit$^{\rm b}$ &  D1 & W3
                 & swarm? \\[0.08cm]
\hline\\[-0.15cm]
C/2010 X11 & $\;\:$5 & C2      & $\pm$0$^\prime\!$.05
           & {\hspace{0.08cm}}$\pm$5$\;\!^\prime\!$.24
           & $\:\,\pm$0$^\prime\!$.81 & No?? \\
C/2010 X12 &      26 & C2,\,C3 & $\pm$0.46 & $\:\:\pm$0.62 & $\:\:\pm$7.08
           & Yes \\
C/2010 X13 & $\;\:$9 & C2      & $\pm$0.29 & $\pm$14.63 & $\pm$20.46 & Yes?? \\
C/2010 X14\rlap{$^{\mbox{\boldmath $\star$}}$} &      49 & C2,\,C3
           & $\pm$0.69 & $\pm$15.58 & $\pm$24.87 & Yes \\
C/2010 X15 &      17 & C2,\,C3 & $\pm$0.52 & $\:\:\pm$5.51 & $\:\:\pm$4.84
           & ? \\
C/2010 X16 &      27 & C2,\,C3 & $\pm$0.47 & $\pm$29.65 & $\pm$38.11 & Yes \\
C/2010 X17 &      18 & C2,\,C3 & $\pm$0.42 & $\:\:\pm$0.56 & $\:\:\pm$8.71
           & Yes \\
C/2010 Y1  &      11 & C2      & $\pm$0.27 & $\:\:\pm$0.79 & $\:\:\pm$7.80
           & Yes \\
C/2010 Y2  &      18 & C2      & $\pm$0.28 & $\pm$14.04 & $\pm$23.02 & Yes \\
C/2010 Y3  &      24 & C2,\,C3 & $\pm$0.80 & $\:\:\pm$1.34 & $\:\:\pm$8.73
           & Yes \\
C/2010 Y4  &      12 & C2      & $\pm$0.40 & $\:\:\pm$3.20 & $\:\:\pm$6.17
           & Yes\\
C/2010 Y5  &      27 & C2,\,C3 & $\pm$0.62 & $\:\:\pm$1.40 & $\pm$10.36 & Yes \\
C/2010 Y6  &      19 & C2      & $\pm$0.42 & $\:\:\pm$3.22 & $\:\:\pm$7.34
           & Yes \\
C/2010 Y7  &      27 & C2,\,C3 & $\pm$0.47 & $\pm$20.77 & $\pm$30.07 & Yes \\
C/2010 Y8  &      16 & C2      & $\pm$0.41 & $\:\:\pm$0.81 & $\pm$10.18 & Yes \\
C/2010 Y9  &      25 & C3      & $\pm$0.44 & $\:\:\pm$8.58 & $\pm$17.64 & Yes \\
C/2010 Y10\rlap{$^{\mbox{\boldmath $\star$}}$} & \llap{1}88
   & C2,\,C3\rlap{$^{\rm c}$} & $\pm$0.53 & $\:\:\pm$5.88 & $\pm$51.64 & Yes \\
C/2010 Y11 &      19 & C2,\,C3 & $\pm$0.69 & $\:\:\pm$7.71 & $\pm$18.42 & Yes \\
C/2010 Y12 &      16 & C2 & $\pm$0.19 & $\:\:\pm$5.65 & $\:\:\pm$6.50 & Yes? \\
C/2010 Y13 &      14 & C2      & $\pm$0.28 & $\:\:\pm$2.73 & $\:\:\pm$9.68
           & Yes \\
C/2010 Y14 &      13 & C2      & $\pm$0.53 & $\:\:\pm$5.76 & $\:\:\pm$5.91
           & ? \\
C/2010 Y15\rlap{$^{\mbox{\boldmath $\star$}}$} &      59 & C2,\,C3
         & $\pm$0.36 & $\:\:\pm$4.10 & $\pm$19.23 & Yes \\[0.03cm]
\hline\\[-0.25cm]
\multicolumn{7}{l}{\parbox{8.25cm}{$^{\rm a}$\,{\scriptsize Objects with
a star{\vspace{-0.05cm}} are those observed very extensively and having the
angular orbital elements, as computed by Gray, rather{\vspace{-0.05cm}}
strongly resembling those of C/1843~D1.}}} \\[-0.05cm]
\multicolumn{7}{l}{\parbox{8.25cm}{$^{\rm b}$\,{\scriptsize Taken from {\tt
http:/{\hspace{-0.05cm}}/www.projectpluto.com/soho/soho.htm}.}}} \\[-0.06cm]
\multicolumn{7}{l}{\parbox{8.25cm}{$^{\rm c}$\,{\scriptsize Plus HI1-A and
HI1-B.}}}\\[-0.4cm]
\end{tabular}}
\end{center}
\end{table}

Gray's (2013) very recent publication of the orbital elements of the
SOHO/STEREO sungrazers from the second half of 2010 includes all 22
objects of the December 2010 clump.  Their definitive designations are
C/2010~X11 through C/2010~X17 and C/2010~Y1 through C/2010~Y15.  Their
astrometric observations, now available as well, serve here to examine
the similarity of their orbits to those of C/1843~D1 and C/2011~W3.  For
each sungrazer, Table 5 provides the number of observations used, the
coronagraphs employed, and the RMS residuals from Gray's best-fit parabolic
solution and from the elliptical orbits for C/1843~D1 (Sekanina \&
Chodas 2008) and C/2011~W3 (Sekanina \& Chodas 2012) in columns~2 to 6,
respectively.  The last column conveys our conclusion whether or not
each of the objects is likely to be associated with C/1843~D1 and
therefore with the swarm of bright dwarf sungrazers.

A cursory glance through Gray's list of orbits suggests at once that up
to five of them have angular orbital elements that resemble those of
C/1843~D1.  They include the three most extensively observed among the
22~sungrazers --- C/2010~Y10, C/2010~Y15, and C/2010~X14, which are in
Table~5 labeled with a star.  The other two sungrazers are C/2010~X16 and
C/2010~Y12.  The number of observations, on which their orbits are based,
are comparable to those for the orbits of C/2010~X12 and C/2010~X17, which
do not resemble that of C/1843~D1, yet the residuals from their forced
solutions in colmn~5 are nearly as good as those from the best-fit
solutions.  Thus, for C/2010~X16 and C/2010~Y12 the orbital agreement
with C/1843~D1 does not offer much confidence in the quality of the
published solutions and a comparison of the residuals in columns~5 and
6 should be the decisive criterion, just as is the case with the rest of
Table~5.

Overall, the mean residuals in columns~5 and 6 suggest that the orbits of
19 of the 22 objects are better, usually much better, fitted by the orbit of
C/1843~D1 than C/2011~W3.  The association with the swarm of bright dwarf
sungrazers is well established for 17, a little less so for C/2010~Y12
(because of the competing residuals in columns~5 and 6), and problematic but
still possible for C/2010~X13 (because of the very small number of used
observations).

We have serious doubts about the direct relationship to the swarm of bright
dwarf sungrazers only for three members of the clump in Table~5:\ C/2010~X11,
C/2010~X15, and C/2010~Y14.  Information on the first object is based on only
five observations, while other two show nearly equal mean residuals from the
two reference orbits; there is thus very little evidence on which a judgment
could be predicated.

As a general comment on column 5 of Table~5, the high mean residuals from
the orbit of comet C/1843~D1 displayed by C/2010~X14, C/2010~X16, C/2010~Y2,
and C/2010~Y7 are a product of these objects' perihelion distances appreciably
exceeding that of C/1843~D1, and should not alarm the reader.  

We conclude that a {\it great majority of the clump of faint Kreutz sungrazers
from the period of December 13--20, 2010 was associated with C/1843~D1 and
therefore with the swarm of bright dwarf sungrazers\/}.

Nothing definite can at present be concluded about the clump of faint
SOHO/STEREO sungrazers from December 2011.  This problem must wait until the
astrometric positions and orbits of these sungrazers become available.

The spike-like features in the distribution of arrivals of faint SOHO/STEREO
Kreutz minicomets have their counterparts among the bright dwarf sungrazers as
well (step-like features; see Sec.~2 and Figure 1).  The four objects brighter
at maximum than magnitude~3 that arrived between 2010.00 and 2010.27
(C/2009~Y4, C/2010~B3, C/2010~E6, and C/2010~G4) fit a slope distinctly steeper
than the polynomial's slope and so do the three objects between 2010.80 and
2010.88 (C/2010~U8, C/2010~V8, and C/2010~W2).  The equivalent rates are,
respectively, $\sim$15 and nearly 40 per year, several times greater than
the peak rate of the smoothed distribution.
\begin{figure*}
\vspace{-14.1cm}
\hspace{-2.01cm}
\centerline{
\scalebox{0.78}{
\includegraphics{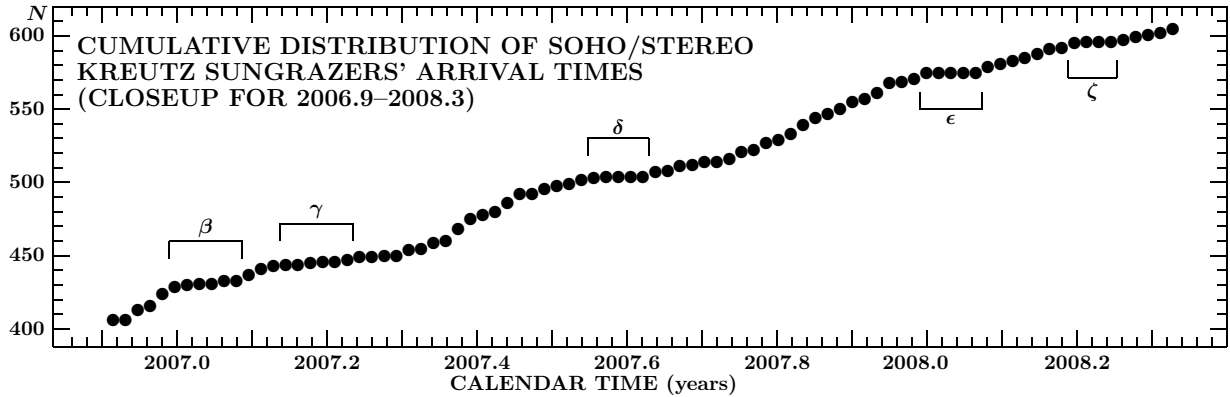}}} 
\vspace{-3.75cm}
\caption{Closeup of the cumulative distribution of arrival times of the
SOHO/STEREO Kreutz sungrazers between 2006.9 and 2008.3.  Compared to
Figure 4, the flat segments $\beta$--$\zeta$ are much better resolved
because in a limited time span the curve is now scaled
differently.{\hspace*{-0.05cm}}{\vspace*{0.45cm}}}
\end{figure*}

We detect a yet another effect that shows that during the few years around
2010 the distribution in Figure 3 behaves untypically.  It is the relation
between the May and the October-December peaks that we already mentioned
in Sec.\ 2.  While the May spikes are never smaller than the October-December
spikes in 2004--2007 and 2012, the former are outperformed by the latter in
2008--2011.

The cumulative distribution of arrival times of all Kreutz sungrazers
seen with the SOHO and STEREO coronagraphs between 2004 and 2013 is
presented in Figure 4, in which we plot the number of objects after each
six-day interval of time.  The averaged slope of the distribution is
practically constant until about the end of 2008, then it picks up a
little, as can be expected from the distribution in Figure 3.

Peculiar features on the cumulative-distribution curve in Figure 4 are
flat segments or ``flats,'' {\it fairly prolonged\/} intervals of time
during which the distribution curve essentially levels off, with a zero
or near-zero arrival rate of the SOHO/STEREO sungrazers.  Their times do not
correlate well with those of SOHO's roll and must be a true property
of the Kreutz population's distribution along the orbit.  Because of the
extremely compressed scale in Figure 4, the flats appear very short.  The
appearance of flats is in greater detail illustrated on a section of the
cumulative distribution in Figure 5.  Each flat was required to satisfy
at least one of two conditions:\ either no sungrazers detected to arrive
for a minimum of three six-day consecutive intervals, or their total number
not exceeding four over a minimum of five six-day consecutive intervals.
Statistically, one would expect the highest rates of flats during 2004--2005
because of lower arrival rates, no high spikes on the distribution curve
in Figure~3, and the absence of bright SOHO/STEREO sungrazers.  However,
Figure 4 shows that this was not the case, suggesting instead a seemingly
counterintuitive positive correlation between the bright dwarf sungrazers'
arrival rate and the rate of flats on the cumulative-distribution curve during
the period ending with 2008:\ there were no flats between the start of 2004
and mid-2006, while during the next two years, between the arrivals of the
bright sungrazers C/2006~A5, C/2006~V2, and C/2008~K4, the average rate was
3 flats per year.  From mid-2009 to mid-2011 the rate of flats was nearly
constant at 2 per year.

\section{Discussion}

The properties of the distribution of arrivals of the SOHO/STEREO
sungrazers in the years 2004--2013 illustrate the morphological complexity
of the Kreutz system, which is driven by --- and can be understood
qualitatively and, to a degree, even quantitatively, in terms of --- a
cascading fragmentation process, proposed by Sekanina (2002) and in
greater detail by Sekanina and Chodas (2007).  In this model, the
fragments continue to break up over and over again at all heliocentric
distances, so that an enormous and {\it essentially\/} continuous stream
of subfragments (or high-generation fragments) arrives at perihelion at
any time.  This stream, which forms a ring along the orbit, is, however,
not structureless, and the degree of its nonuniformity varies from place
to place on many spatial scales that are equivalent to time scales from
days to dozens of years and longer.  The segment of the Kreutz system
that arrived at perihelion in 2004--2005 was apparently characterized by
a high degree of uniformity, while in the subsequent years the opposite
was the case.  

The existence of brighter objects (\mbox{A--U} in Figures 3 and 4) in
the cloud of SOHO/STEREO minicomets suggests that, among the sungrazers,
they may consist of a material of a greater than average cohesion, so that
their fragmentation rate is slower than that of other fragments in the
stream.  Because their parent(s) broke up less often and probably also at
later times than the rest, the spatial distribution of subfragments that
derive from such objects is less uniform and more cluster-like, which
necessarily leads to greater variations and relatively often to {\it gaps\/}
in their distribution along the orbit.  These gaps show up as the flat
segments on the cumulative-distribution curve.  Thus, to a point, the
occurrence rate of flats should indeed vary in some proportion to the
arrival rate of bright dwarf fragments.  Only when this rate increases
significantly is there much more debris around, so that the number of flats
begins to drop, but even then they should not be entirely absent.

The orbital results for the bright dwarf sungrazers in Table 2, which
completely rule out any potential relationship between their sharply elevated
rate and comet C/2011~W3, are supported by the evidence of very few {\it
faint\/} SOHO/STEREO Kreutz minicomets in orbits similar to that of C/2011~W3.
Table~6 shows that there were only nine such objects in the set of more than
1000 SOHO/STEREO Kreutz sungrazers between 2004 and 2013 with currently known
orbits.  Faint companions in close proximity of C/2011~W3 were recent
fragments whose existence does not change the overall picture.
\begin{table*}[ht]
\noindent
\begin{center}
{\footnotesize {\bf Table 6}\\[0.08cm]
{\sc List of Orbital Elements and Apsidal Orientation for SOHO/STEREO
Sungrazers with Available Orbits in 2004--2013\\Moving in Paths Generally
Similar to That of C/2011 W3 (Eq.\ J2000.0).}\\[0.08cm]
\begin{tabular}{l@{\hspace{0.4cm}}l@{\hspace{0.07cm}}c@{\hspace{0.07cm}}r@{\hspace{0.43cm}}c@{\hspace{0.38cm}}c@{\hspace{0.38cm}}c@{\hspace{0.26cm}}c@{\hspace{0.35cm}}c@{\hspace{0.35cm}}c@{\hspace{0.35cm}}c@{\hspace{0.15cm}}c@{\hspace{0.18cm}}c@{\hspace{0.58cm}}c}
\hline\hline\\[-0.24cm]
 & \multicolumn{7}{@{\hspace{-0.17cm}}c}{Orbital elements}
 & \multicolumn{2}{@{\hspace{-0.1cm}}c}{Line of apsides} & & Number &
 & \\[-0.07cm]
 & \multicolumn{7}{@{\hspace{-0.17cm}}c}{\rule[0.6ex]{7cm}{0.4pt}}
 & \multicolumn{2}{@{\hspace{-0.1cm}}c}{\rule[0.6ex]{2.2cm}{0.4pt}}
 & Corona- & of used & & \\[-0.07cm]
Object
 & \multicolumn{3}{@{\hspace{-0.1cm}}c}{$t_\pi$ (ET)}
 & $\omega$ & $\Omega$ & $i$ & $q$\,($R_{\mbox{\boldmath $\scriptstyle
   \odot$}}$) & $L_\pi$ & $B_\pi$ & graph(s) & positions & Author
 & Reference \\[0.05cm]
\hline\\[-0.22cm]
C/2004 A4 & 2004 & Jan. &  8.64 & 60$^\circ\!\!$.94 & 333$^\circ\!\!$.15
          & 134$^\circ\!\!$.78 & 1.03 & 281$^\circ\!\!$.42
          & +38$^\circ\!\!$.35 & C3 & 15 & Marsden & MPC\,60097 \\
C/2004 P5 & 2004 & Aug. &  9.45 & 52.72 & 324.88 & 133.10 & 1.31 & 282.97
          & +35.52 & C3 & 25 & Marsden & MPC\,52766 \\
C/2005 U5 & 2005 & Oct. & 22.20 & 49.13 & 319.04 & 131.09 & 1.05 & 281.82
          & +34.75 & C3 & 15 & Marsden & MPC\,55718 \\
C/2006 U8 & 2006 & Oct. & 17.09 & 53.25 & 319.52 & 137.33 & 1.07 & 274.96
          & +32.89 & C2,\,C3 & 10 & Marsden & MPC\,58538 \\
C/2007 S4 & 2007 & Sept. & 24.31 & 57.15 & 325.85 & 138.08 & 1.07 & 276.80
          & +34.14 & C2,\,C3 & 14 & Marsden & MPC\,60927 \\
C/2010 C4 & 2010 & Feb. &  7.99 & 61.46 & 333.75 & 135.99 & 1.10 & 280.85
          & +37.62 & C3 & 14 & Marsden & MPC\,72131 \\
C/2010 Q2 & 2010 & Aug. & 24.79 & 51.38 & 325.82 & 133.48 & 1.73 & 285.08
          & +34.54 & C2,\,C3 & 62 & Gray & MPC\,84618 \\
C/2010 Y16 & 2010 & Dec. & 21.83 & 57.04 & 329.47 & 133.49 & 1.67 & 282.76
           & +37.50 & C3 & 15 & Gray & MPC\,84623 \\
C/2013 F4 & 2013 & Mar. & 27.51 & 60.71 & 336.38 & 138.47 & 1.79 & 283.22
          & +35.33 & C3\rlap{$^{\rm a}$} & \llap{1}26 & Gray
          & MPC\,84626\\[0.05cm]
\hline\\[-0.27cm]
\multicolumn{14}{l}{\parbox{12cm}{$^{\rm a}$\,\scriptsize Also:\ COR2-A,
 COR2-B, and HI1-B.}}\\[0.05cm]
\end{tabular}}
\end{center}
\end{table*}

{\it What are these conflicting lines of evidence telling us?}  By far the
most plausible hypothesis is that {\it another Lovejoy-sized fragment\/},
closely related to C/1843~D1 and moving in a similar orbit but with its
perihelion time shortly preceding that of C/2011~W3, {\it continued to
break up during its orbital motion still at large heliocentric distances\/}.
This hypothesis logically explains several major findings of our paper.

First, the {\it 16~tightly related bright dwarf sungrazers in the swarm\/},
identified in Sec.~3 and Table~2, are the {\it prime fragmentation
products of the postulated object --- their parent\/}.  Indeed, adding up
the brightness contributions from the 16~sungrazers in Table 1 leads to an
equivalent of apparent magnitude $-$2.0, not much fainter than was C/2011~W3
(whose apparent magnitude at maximum light was estimated at $-$3; e.g.,
Green 2011).

Second, a population of fainter dwarf sungrazers must be, in addition to the
16~bright ones, also derived from the same parent.  An example of direct
evidence for this argument is the {\it clump\/} of 22 faint objects from
December 2010, most of which were shown to be associated with the swarm of
bright dwarf sungrazers, representing in this scenario a {\it debris of a
sizable, but less cohesive member of this swarm that fragmented long before
reaching perihelion\/}.  This clump is probably the most extreme example
of this advanced stage of cascading fragmentation but certainly not the
only one.

Third, a {\it hint of a hierarchy in the distribution of fragment dimensions
is perceived in both the swarm and the clump\/}.  The swarm's brightest object,
SOHO-2143, is in about the same position relative to the apparent ends of
the swarm modeled in Figure 2, as is the clump's most prominent object,
C/2010~Y10, relative to the apparent ends of the clump.  In both cases this
``primary'' fragment is in the feature's second half.  Its positioning may
provide information on the vector field of separation velocities and on the
fragmentation mechanism's nature.

Fourth, this hypothesis is consistent with our explanation of the {\it gaps\/}
in the distribution of arrival times of the SOHO/STEREO Kreutz population
or, equivalently, the {\it flats\/} in its cumulative distribution, as
described in Sec.~4 and earlier in this section.

Fifth, the presented hypothesis explains the {\it fairly well-defined limits
of the tightly-knit swarm\/}.  It is in fact remarkable that SOHO-2571, the last
point on the cumulative distribution of bright dwarf sungrazers in Figure~1,
has orbital elements so dramatically different from those of the swarm's
members.  The future will show whether this comet is a ``wanderer'' (like
bright SOHO sungrazers before 2004) or the first member of a new swarm.

\begin{table}[b]
\vspace*{-0.01cm}
\noindent
\begin{center}
{\footnotesize {\bf Table 7}\\[0.08cm]
{\sc Perturbation of Arrival Time (yr) at Next Perihelion by
Separation Velocity Acquired in Fragmentation Event\\as Function of
Heliocentric Distance.}\\[0.08cm]
%
\begin{tabular}{c@{\hspace{0.1cm}}c@{\hspace{0.3cm}}c@{\hspace{0.25cm}}c@{\hspace{0.45cm}}c@{\hspace{0.25cm}}c}
\hline\hline\\[-0.28cm]
Distance & Elapsed & \multicolumn{2}{@{\hspace{-0.15cm}}c}{Radial velocity}
 & \multicolumn{2}{@{\hspace{-0.05cm}}c}{Transverse velocity} \\[-0.06cm]
from Sun$^{\rm a}$ & time$^{\rm b}$
 & \multicolumn{2}{@{\hspace{-0.15cm}}c}{\rule[0.8ex]{2.6cm}{0.4pt}}
 & \multicolumn{2}{@{\hspace{-0.05cm}}c}{\rule[0.8ex]{2.6cm}{0.4pt}}\\[-0.06cm]
(AU) & (yr) & \raisebox{0.4ex}{\scriptsize $\:$+1ms$^{-1}$}
            & \raisebox{0.4ex}{\scriptsize $\:-$1ms$^{-1}$}
            & \raisebox{0.4ex}{\scriptsize $\:$+1ms$^{-1}$}
            & \raisebox{0.4ex}{\scriptsize $\:-$1ms$^{-1}$} \\[0.05cm]
\hline\\[-0.26cm]
10         &    $\;\:$2.42 & +3.6715 & $-$3.6463 & +0.0891 & $-$0.0888 \\
20         &    $\;\:$6.96 & +2.5195 & $-$2.5074 & +0.0446 & $-$0.0443 \\
30         &         13.01 & +1.9941 & $-$1.9865 & +0.0297 & $-$0.0294 \\
50         &         29.10 & +1.4441 & $-$1.4400 & +0.0178 & $-$0.0175 \\
75         &         56.47 & +1.0707 & $-$1.0683 & +0.0117 & $-$0.0115 \\
\llap{1}00 &         92.79 & +0.8261 & $-$0.8246 & +0.0086 & $-$0.0083 \\
\llap{1}30 & \llap{1}51.88 & +0.6061 & $-$0.6052 & +0.0063 & $-$0.0060 \\
\llap{1}60 & \llap{2}40.25 & +0.4178 & $-$0.4174 & +0.0045 & $-$0.0043 \\
\llap{1}86\rlap{.43}
           & \llap{4}50.00 & +0.1851 & $-$0.1849 & +0.0025 & $-$0.0023 \\
\llap{1}60 & \llap{6}59.75 & +0.0649 & $-$0.0648 & +0.0012 & $-$0.0011 \\
\llap{1}30 & \llap{7}48.12 & +0.0325 & $-$0.0324 & +0.0007 & $-$0.0007 \\
\llap{1}00 & \llap{8}07.21 & +0.0159 & $-$0.0159 & +0.0004 & $-$0.0004 \\
        75 & \llap{8}43.53 & +0.0079 & $-$0.0079 & +0.0002 & $-$0.0002 \\
        50 & \llap{8}70.90 & +0.0032 & $-$0.0032 & +0.0001 & $-$0.0001 \\
        30 & \llap{8}86.99 & +0.0011 & $-$0.0011 & +0.0001 & $-$0.0001 \\
        20 & \llap{8}93.04 & +0.0005 & $-$0.0005 & $\;\;\:$0.0000
           & $\;\;\:$0.0000 \\
        10 & \llap{8}97.58 & +0.0001 & $-$0.0001 & $\;\;\:$0.0000
           & $\;\;\:$0.0000 \\[0.05cm]
\hline \\[-0.3cm]
\multicolumn{6}{l}{\parbox{8cm}{$^{\rm a}$\,{\scriptsize At the time of
separation; aphelion is at 186.43 AU.}}} \\[-0.12cm]
\multicolumn{6}{l}{\parbox{8.0cm}{$^{\rm b}$\,{\scriptsize Measured from
 previous perihelion for orbital period of 900 yr.}}}\\[-0.2cm]
\end{tabular}}
\end{center}
\end{table}

More broadly, the hypothesis of a broken-up parent sungrazer is closely
linked to, and symptomatic of, the process of nucleus fragmentation at large
heliocentric distance, whose important trait is its fairly insignificant
effect on the orbital periods of the parent's fragments (Sekanina 2002).
Accordingly, they arrive at perihelion nearly simultaneously (on a scale of
the orbital period) as a {\it true swarm\/}.  For illustration, Table 7
lists the perturbations of the arrival time at next perihelion for a fragment
that separated far from the Sun with a typical velocity of 1~m~s$^{-1}$ in
either of two cardinal directions in the plane of a sungrazing orbit with a
peri\-helion distance of 1.2~{\Rsun} and an orbital period of 900~years
(with aphelion at 186.43~AU).  The radial component, $V_R$, is directed away
from (+) or toward the Sun, the transverse component, $V_T$, is in the orbit
plane and perpendicular to $V_R$, pointing in the general direction of the
orbital motion (+) or in the opposite direction.  There is no contribution
from the normal component.

Table 7 shows some noteworthy facts.  First of all, the radial separation
velocity is 10 to 100 times more efficient in triggering off the same
change in the arrival time at next perihelion passage than is the transverse
velocity.  Second, the effect is approximately, but not exactly, symmetrical
for positive and negative velocities.  And third, a peak arrival rate of 4.6
per year for the bright dwarf sungrazers (Sec.~2) is equivalent to their
minimum temporal separation of 0.22 yr, which in Table 7 matches the
arrival-time perturbation for a separation near aphelion.  If the separation
velocity should be lower than 1~m~s$^{-1}$, the assumed fragmentation event
would have taken place before aphelion, and vice versa.  The temporal relations
among the bright dwarf sungrazers from \mbox{2006--2013} are therefore on the
right order of magnitude to be consistent with the process of cascading
fragmentation.

\begin{table}[t]
\noindent
\begin{center}
{\footnotesize {\bf Table 8}\\[0.08cm]
{\sc Perturbations of Argument of Perihelion ($\omega$),\\Longitude of
Ascending Node ($\Omega$), and Inclination ($i$)\\by Normal Component of
Separation Velocity\\Acquired in Fragmentation Event as Function\\of
Heliocentric Distance.}\\[0.08cm]
\begin{tabular}{c@{\hspace{0.08cm}}c@{\hspace{0cm}}c@{\hspace{0.11cm}}c@{\hspace{0cm}}c@{\hspace{0.11cm}}c@{\hspace{0cm}}c}
\hline\hline\\[-0.28cm]
Distance & \multicolumn{2}{@{\hspace{0.05cm}}c}{Effect in $\omega$}
         & \multicolumn{2}{@{\hspace{0.05cm}}c}{Effect in $\Omega$}
         & \multicolumn{2}{@{\hspace{0.05cm}}c}{Effect in $i$} \\[-0.06cm]
from Sun$^{\rm a}$
 & \multicolumn{2}{@{\hspace{0.05cm}}c}{\rule[0.8ex]{2.05cm}{0.4pt}}
 & \multicolumn{2}{@{\hspace{0.05cm}}c}{\rule[0.8ex]{2.05cm}{0.4pt}}
 & \multicolumn{2}{@{\hspace{0.05cm}}c}{\rule[0.8ex]{2.05cm}{0.4pt}} \\[-0.06cm]
(AU) & \raisebox{0.4ex}{\scriptsize $\:$+1ms$^{-1}$}
     & \raisebox{0.4ex}{\scriptsize $\:-$1ms$^{-1}$}
     & \raisebox{0.4ex}{\scriptsize $\:$+1ms$^{-1}$}
     & \raisebox{0.4ex}{\scriptsize $\:-$1ms$^{-1}$}
     & \raisebox{0.4ex}{\scriptsize $\:$+1ms$^{-1}$}
     & \raisebox{0.4ex}{\scriptsize $\:-$1ms$^{-1}$} \\[0.05cm]
\hline\\[-0.22cm]
10         & $-$0$^\circ\!\!$.25 & +0$^\circ\!\!$.25 & $-$0$^\circ\!\!$.31
           & +0$^\circ\!\!$.31 & $-$0$^\circ\!\!$.03 & +0$^\circ\!\!$.03 \\
20         & $-$0.50 & +0.50 & $-$0.62 & +0.62 & $-$0.06 & +0.06 \\
30         & $-$0.75 & +0.75 & $-$0.92 & +0.93 & $-$0.09 & +0.08 \\
50         & $-$1.25 & +1.26 & $-$1.54 & +1.55 & $-$0.14 & +0.12 \\
75         & $-$1.87 & +1.90 & $-$2.31 & +2.33 & $-$0.21 & +0.17 \\
\llap{1}00 & $-$2.49 & +2.53 & $-$3.07 & +3.11 & $-$0.29 & +0.21 \\
\llap{1}30 & $-$3.23 & +3.30 & $-$3.99 & +4.05 & $-$0.38 & +0.25 \\
\llap{1}60 & $-$3.97 & +4.06 & $-$4.90 & +4.99 & $-$0.48 & +0.28 \\
\llap{1}86\rlap{.43}
           & $-$4.61 & +4.73 & $-$5.69 & +5.81 & $-$0.56 & +0.29 \\
\llap{1}60 & $-$3.98 & +4.06 & $-$4.91 & +4.99 & $-$0.45 & +0.25 \\
\llap{1}30 & $-$3.24 & +3.30 & $-$4.00 & +4.05 & $-$0.35 & +0.21 \\
\llap{1}00 & $-$2.50 & +2.53 & $-$3.08 & +3.11 & $-0.25$ & +0.17 \\
        75 & $-$1.88 & +1.90 & $-$2.32 & +2.33 & $-$0.18 & +0.13 \\
        50 & $-$1.26 & +1.27 & $-$1.55 & +1.56 & $-$0.11 & +0.09 \\
        30 & $-$0.76 & +0.76 & $-$0.93 & +0.93 & $-$0.06 & +0.05 \\
        20 & $-$0.50 & +0.51 & $-$0.62 & +0.62 & $-$0.04 & +0.03 \\
        10 & $-$0.25 & +0.25 & $-$0.31 & +0.31 & $-$0.02 & +0.01 \\[0.05cm]
\hline \\[-0.25cm]
\multicolumn{7}{l}{\parbox{7.8cm}{$^{\rm a}$\,{\scriptsize At the time of
separation; aphelion is at 186.43 AU.}}} \\[-0.2cm]
\end{tabular}}
\end{center}
\end{table}

The argument of perihelion $\omega$ and the longitude of the ascending node
$\Omega$ are both affected by fragmentation at large heliocentric distance
rather strongly, while the inclination $i$ to a much lesser degree, as is
shown in Table~8.  It is now the normal component, $V_N$, of the separation
velocity that is primarily involved, pointing to the north orbital pole (+) or
in the opposite direction.  There is no contribution from the radial component
and, only in $\omega$, a very minor one, not exceeding 0$^\circ\!$.02 per
1~m~s$^{-1}$ velocity from~the transverse component.  The normal-velocity
effect is again nearly symmetrical relative to aphelion and also for
positive and negative velocities.  An exception is the inclination, with the
negative velocity having a generally smaller effect, as small at aphelion as
one half of that of the positive velocity of the same magnitude.
\begin{table}[b]
\noindent
\begin{center}
{\footnotesize {\bf Table 9}\\[0.08cm]
{\sc Perturbation of Perihelion Distance ({\Rsun}) by
 Transverse\\[-0.03cm]Component of Separation Velocity Acquired
 in\\Fragmentation Event as Function\\of
Heliocentric Distance.}\\[0.1cm]
\begin{tabular}{c@{\hspace{0.15cm}}c@{\hspace{0cm}}c@{\hspace{0.1cm}}c@{\hspace{0.15cm}}c@{\hspace{0.2cm}}c@{\hspace{0cm}}c}
\hline\hline\\[-0.35cm]
 & & & \vline & & & \\[-0.25cm]
Distance & \multicolumn{2}{@{\hspace{0cm}}c}{Transverse velocity}
 & \vline & Distance
 & \multicolumn{2}{@{\hspace{0cm}}c}{Transverse velocity} \\[-0.06cm]
from Sun & \multicolumn{2}{@{\hspace{0cm}}c}{\rule[0.8ex]{2.5cm}{0.4pt}}
 & \vline & from Sun$^{\rm a}$
 & \multicolumn{2}{@{\hspace{0cm}}c}{\rule[0.8ex]{2.5cm}{0.4pt}}\\[-0.06cm]
(AU) & \raisebox{0.4ex}{\scriptsize $\;$+1\,m\,s$^{-1}$}
     & \raisebox{0.4ex}{\scriptsize $\;-$1\,m\,s$^{-1}$}
     & \vline
     & (AU) & \raisebox{0.4ex}{\scriptsize $\;$+1\,m\,s$^{-1}$}
     & \raisebox{0.4ex}{\scriptsize $\;-$1\,m\,s$^{-1}$} \\[0.05cm]
\hline\\[-0.4cm]
           &        &          & \vline &               &        & \\[-0.15cm]
10         & +0.008 & $-$0.008 & \vline & 100           & +0.077 & $-$0.075 \\
20         & +0.015 & $-$0.015 & \vline & 130           & +0.101 & $-$0.097 \\
30         & +0.023 & $-$0.023 & \vline & 160           & +0.125 & $-$0.119 \\
50         & +0.038 & $-$0.038 & \vline & 186\rlap{.43} & +0.146 & $-$0.138 \\
75         & +0.058 & $-$0.057 & \vline &               & \\[0.05cm]
\hline \\[-0.3cm]
\multicolumn{7}{l}{\parbox{7cm}{$^{\rm a}$\,{\scriptsize At the time of
separation; aphelion is at 186.43 AU.{\vspace*{-0.2cm}}}}}
\end{tabular}}
\end{center}
\end{table}

The perturbations of the three angular elements, which always have the same
sign, are only very approximately in constant proportions at different
heliocentric distances.  At aphelion, for example, the
\mbox{$\omega$:$\Omega$:$i$} perturbation ratios are
\mbox{0.81:1:(0.05-0.10)}.  Given the errors involved, these
ratios are in good agreement (including the correct signs) with the
relative ratios of \mbox{0.88:1:0.024} of the average slopes in Table~3.
This match is a very encouraging indication that the angular-element
trends among the tightly-knit swarm's sungrazers were indeed triggered by
perturbations due to separation velocities acquired in fragmentation
events involving these objects.  The ratios also explain why in Table~2
the trends in $\omega$ and $\Omega$ are obvious but those in $i$ are not.  

Fragmentation always has an extremely small effect on the orientation of
the line of apsides.  The magnitude of these perturbations never exceeds
0$^\circ\!$.03 per 1~m~s$^{-1}$ at $>$10~AU from the Sun and they are
caused by the transverse component of the separation velocity in the
perihelion longitude $L_\pi$ and by the normal component in the perihelion
latitude $B_\pi$.

The perturbations of the perihelion distance for a fragment that broke
off from its parent far from the Sun are dominated by the transverse
component of the separation velocity, as shown in Table 9.  There is
a perfect symmetry relative to aphelion, so only the preaphelion
numbers are listed.  On the other hand, the effect's symmetry due to a
positive (in the general direction of the orbital motion) vs a negative
velocity is only approximate.  There is no contribution from the radial
component and only a minor contribution, not exceeding 0.004~{\Rsun} per
1~m~s$^{-1}$, from the normal component. It is noted that even a
submeter-per-second separation velocity acquired during a {\it single\/}
fragmentation event near aphelion can turn a fragment's sungrazing orbit
with a perihelion distance of 1.1~{\Rsun} into a sun-striking orbit.  It
is noted that two objects in Table 2 do indeed have their perihelion
distances smaller than 1~{\Rsun}.

Finally, perturbations of the eccentricity due to fragmentation at large
heliocentric distance are not discussed here in any detail because of
the parabolic orbital approximation used for the swarm members.  We only
note that they are expressible in terms of the perturbations of the
perihelion distance and the semimajor axis, whose reciprocal, $1/a$, is
affected by less than 0.00003~AU$^{-1}$ at $>$10~AU from the Sun primarily
by the radial component.  Unlike near the Sun, the transverse component
is much less important and there is no contribution to $1/a$ from the
normal component.

To summarize up to this point, our discussion suggests that the orbital
perturbations of fragments due to submeter- or meter-per-second separation
velocities acquired during breakup events at large heliocentric distances
are generally consistent with the systematic trends in the orbital elements
in Table 2, supporting our hypothesis that the tightly-knit swarm of
16~bright dwarf sungrazers (as well as a debris of additional, fragmented
members of the swarm), peaking in 2010/2011, consisted of fragments of an
originally common parent, which itself must have been a fragment of a
precursor that also gave birth to comet C/1843~D1.  This precursor could
have been the celebrated comet X/1106~C1, as suggested by Sekanina \& Chodas
(2007).  The arrival time of the swarm, about 168 years after the appearance
of C/1843~D1, requires a relative separation velocity of only about
1~m~s$^{-1}$ between the swarm's parent and this 19th-century sungrazer,
if they detached from X/1106~C1 right at perihelion, or somewhat greater
than 1~m~s$^{-1}$, if a little off the perihelion point, but in any case
a velocity typical for separations of sungrazer fragments near perihelion
(Sekanina \& Chodas 2007).  It is possible that C/1843~D1, the swarm's
parent, many additional sizable sungrazers as well as huge amounts of debris
had separated from X/1106~C1 as a single object, which became subject
to sudden or gradual fragmentation only during its liftoff through the
precursor's atmosphere and/or soon afterwards.

Because SOHO-2143 was clearly the brightest among the swarm's members
(Table 1), it is plausible to deem this sungrazer the most massive fragment,
located near the position that would have been occupied by the parent had
it survived intact.  The swarm's 15 remaining objects are then products of
one or more breakup events, in harmony with the conceptual paradigm of
cascading fragmentation (Sekanina 2002, Sekanina \& Chodas 2007).  If
the mechanism that accounts for the separation velocities causes them
to prefer a certain direction --- such as, for example, the parent's spin
--- the contributions from the individual fragmentation events to the
 total perturbation effect would simply add up in some direction in an
inertial coordinate system, to which the introduced coordinate system
is, due to the orbit's extreme elongation, practically equivalent at any
point far enough from perihelion.  The task is now reduced to finding
out whether it is at all possible to choose a limited number of, and
appropriate locations for, the breakup events in which a swarm's fragment
could have been reasonably involved, such that the respective orbital
difference from Table~2 is in each of the five elements matched closely
enough by the sum of perturbations from these events.

\begin{table*}[t]
\noindent
\begin{center}
{\footnotesize {\bf Table 10}\\[0.08cm]
{\sc Example of Interpreting Orbital Difference Between Two Swarm Members
in Terms of Perturbations\\Acquired During Series of Six Fragmentation
Events.$^{\rm a}$}\\[0.08cm]
\begin{tabular}{l@{\hspace{0.5cm}}c@{\hspace{0.5cm}}c@{\hspace{0.5cm}}c@{\hspace{0.5cm}}c@{\hspace{0.5cm}}c@{\hspace{0.5cm}}c@{\hspace{0.5cm}}c@{\hspace{0.2cm}}c}
\hline\hline\\[-0.28cm]
Fragmentation event & \#1 & \#2 & \#3 & \#4 & \#5 & \#6 & Sum & Difference \\
%
%
Time from previous perihelion (yr) & $\;\:$2.4 & $\;\:$9.8 & 29.1 & 69.8
                        & 151.9 & \llap{4}50.0 & from & C/2006 A5 \\
Heliocentric distance (AU)         & 10.0 & 25.0 & 50.0 & 85.0 & 130.0
                                   & \llap{1}86.4\rlap{3} & the six & minus \\
Perturbations: & & & & & & & events & SOHO-2143 \\[0.05cm]
%
%
\hline\\[-0.22cm]
Arrival time at next perihelion (yr) & $-$2.32 & $-$1.40 & $-$0.91 & $-$0.61
                                     & $-$0.38 & $-$0.12 & $-$5.74& $-$5.74 \\
Argument of perihelion               & $-$0$^\circ\!$.11 & $-$0$^\circ\!$.26
                 & $-$0$^\circ\!$.51 & $-$0$^\circ\!$.86 & $-$1$^\circ\!$.30
                 & $-$1$^\circ\!$.85 & $-$4$^\circ\!$.89 & $-$5$^\circ\!$.00 \\
Longitude of ascending node          & $-$0$^\circ\!$.13 & $-$0$^\circ\!$.31
                 & $-$0$^\circ\!$.62 & $-$1$^\circ\!$.05 & $-$1$^\circ\!$.60
                 & $-$2$^\circ\!$.28 & $-$5$^\circ\!$.99 & $-$6$^\circ\!$.08 \\
Inclination                          & $-$0$^\circ\!$.01 & $-$0$^\circ\!$.03
                 & $-$0$^\circ\!$.05 & $-$0$^\circ\!$.09 & $-$0$^\circ\!$.15
                 & $-$0$^\circ\!$.22 & $-$0$^\circ\!$.55 & $-$0$^\circ\!$.43 \\
Perihelion distance ({\Rsun})        & $-$0.005 & $-$0.011 & $-$0.023
                & $-$0.039 & $-$0.059 & $-$0.083 & $-$0.220 & $-$0.22 \\[0.05cm]
\hline \\[-0.22cm]
\multicolumn{9}{l}{\parbox{16.3cm}{$^{\rm a}$\,{\scriptsize Best fit under
adopted constraints requires{\vspace{-0.04cm}} that the separation velocity
0.96 m\,s$^{-1}$, with the components (in m\,s$^{-1}$): \mbox{$V_R = -0.62$},
\mbox{$V_T = -0.61$}, and \mbox{$V_N = +0.40$}.  The orbital period is, as
before, assumed to equal 900 years.}}}\\[0.24cm]
\end{tabular}}
\end{center}
\end{table*}

There exists far too little information to derive the unique solution.
We offer one example of a greater than average difficulty to outline the
issues involved and to illustrate the plausibility of our swarm's
fragmentation hypothesis.  The first issues are whether in our scenario
we should (a) presume more than one original parent and (b) consider more
than a single orbit about the Sun.  The answer is no on both counts.  One,
the lifetimes of all SOHO/STEREO Kreutz sungrazers observed to disintegrate
just before perihelion are shorter than one revolution, as otherwise
these objects would have disintegrated just before their previous perihelion
passage and could not be seen nowadays.  And two, even if fragmentation
began before previous perihelion, the secondary parents, which would have
to have been large enough objects to survive, and their own subsequently
generated swarms would be separated from each other by such huge gaps in
time that their association could not easily be recognized.  Thus, our task
is indeed confined to one parent and a single revolution about the Sun.

With these constraints in mind, let us consider the first entry in Table~2,
C/2006~A5.  Its orbital elements suggest that it was a member of the
tightly-knit swarm.  Because it preceded SOHO-2143 (a proxy for the parent)
by as much as 5.74 years, it represents a highly challenging case.  In
concert with our objectives, we ask:\ Could the early arrival of C/2006~A5
and the deviation of its orbital elements from those of SOHO-2143 be
explained by perturbations due to separation velocities acquired in a
sequence of fragmentation events?

In our scenario, the difference of almost 6 years is an {\it integrated\/}
perturbation of the arrival time at next perihelion, that is, the sum
of perturbations at $\nu$ fragmentation events in which this sungrazer
was involved either as part of a larger object in each of the first
$\nu-1$ events or as a stand-alone fragment produced by the last,
$\nu$-th event.

To estimate the number of fragmentation events, we use a simple model,
with the magnitude at maximum light (normalized, strictly, to 1~AU
from Earth, but differing very little from the apparent magnitude in
close proximity of the Sun) as a proxy for the initial mass, since
both quantities were shown to have been tightly correlated (Sekanina
2003).  For C/2006~A5 the magnitude at maximum light was 2.9 (Table 1);
for the parent we approximate the lower limit to its peak brightness by
that of SOHO-2143, that is, by magnitude $-$0.5.  The upper limit to
the peak brightness of the original parent is uncertain.  The object may
have been as bright as magnitude $-$2.0 (see above) or even brighter;
we adopt, rather arbitrarily, magnitude $-$2.5.  If the parent and its
fragments are assumed to have split, in each step of the process, into
two parts of equal mass (50 percent of their immediate parent's mass), the
mass of any fragment after $\nu$ events is a \mbox{2$^{-\nu}$-th} part
of the original parent.  Because the estimated mass of C/2006~A5
comes{\vspace{-0.07cm}} out on our assumptions to be{\vspace{-0.04cm}}
between \mbox{10$^{0.4 \times (2.9+0.5)} \simeq 23$} and \mbox{10$^{0.4
\times (2.9+2.5)} \simeq 145$} times smaller than the original parent's
mass, we obtain a condition \mbox{$4.5 < \nu < 7.2$}.  To the extent
that this scenario applies, C/2006~A5 should have been involved in
\mbox{5--7}~fragmentation events during the 900 years, not counting the
initial breakup of X/1106~C1 near perihelion.

In our example, we deem C/2006 A5 a final product of six fragmentation
events and require that the magnitude of the separation velocity acquired by
it during each event not exceed 1~m~s$^{-1}$.  Since this sungrazer preceded
SOHO-2143, the parent's proxy, the radial component, which dominates
the perturbations of the arrival time at next perihelion, must have been
directed toward the Sun.  We also require, rather arbitrarily, that the
separation velocity vector not vary from event to event and that the last
event have occurred at aphelion.  With these constraints, all that needs
to be chosen in Table~7 (or its expanded version) is six entries such that
their sum is greater in absolute value than 5.74~years.  One of many
solutions that fits well the differences between the orbital elements of
C/2006~A5 and SOHO-2143 in Table~2 is presented in Table~10.

The choice of orbital locations for the fragmentation events is, in
the absence of conditions constraining the fragment's history,
rather arbitrary.  Whether the fragmentation events were a corollary
of major thermal stresses that continued to ravage the body's interior
after perihelion (Sekanina \& Chodas 2012) or more or less spontaneous,
the example in Table~10 conforms to an expectation that they became
gradually less frequent and eventually ceased.  Other scenarios are of
course possible, but, as seen from Table~7, postaphelion fragmentation
fits only differences of $<$1 year or so in the arrival times at
next perihelion, unless $|V_N| \gg 1$\,m\,s$^{-1}$ or the number of
fragmentation events is extraordinarily large.

The separation velocity that fits the fragmentation sequence in Table~10
is 0.96~m~s$^{-1}$ per event, with the components \mbox{$V_R = -0.62$},
\mbox{$V_T = -0.61$}, and \mbox{$V_N = +0.40$}, all in m~s$^{-1}$.  The
perfect agreement between the perturbation sum and the orbital difference
is to be expected for the arrival time and the perihelion distance, because
the perturbations of these elements are dominated by a single component
of the separation velocity, the radial and the transverse one, respectively.
The agreement between the perturbation sums and the orbital differences in
the three angular elements is another matter, because the perturbations
of all three are dominated by the normal component of the separation
velocity.  Also, when the radial and transverse components exceed in
magnitude the normal component, the contributions from the transverse
velocity to the arrival-time perturbation and from the normal component
to the perihelion-distance perturbation are not entirely negligible
and they impact the balance of the result.  The residuals in Table~10 of
about 0$^\circ\!$.1 in each of the three angular elements suggest that
they are within the errors of observation, thus acceding to the plausibility
of our perturbation scenario and to the status of the bright dwarf sungrazers
in the swarm as fragments of a common parent, sharing orbits very similar
to that of comet C/1843~D1 and being closely related to this spectacular
object.  In the light of this compelling evidence, the remarkable
coincidence in timing between the arrival of the swarm and the appearance
of comet C/2011~W3 is found to be purely fortuitous.

The clump of nearly two dozen faint SOHO/STEREO Kreutz minicomets from
mid-December 2010 is shown to be a debris of the swarm's another member
that fragmented far from the Sun on its way to perihelion.  For the
clump's duration of about eight days, Table~7 suggests that the parent's
fragmentation most probably began more recently than some 150 years ago
and less than $\sim$130~AU from the Sun after aphelion.  When orbits of all
SOHO/STEREO sungrazers from 2011 are available, the origin of the second
major clump, from December of that year, can likewise be examined.

The situation with the swarm of bright dwarf sungrazers is reminiscent of
that with the temporal distribution of the Solar Maximum Mission (SMM)
sungrazers, whose orbits were studied extensively by Marsden (1989).  He
found that, like the swarm members in our investigation, the objects
belonged to what he called {\it Subgroup~I\/}, which includes C/1843~D1,
and that two pairs of them arrived only 12--13 days apart.  Although
the SMM sungrazers' orbits were determined less accurately than those
of the brighter SOHO/STEREO sungrazers, a general tendency is apparent
among the SMM objects' orbits (Marsden 1989, Marsden \& Williams 2008)
for the argument of perihelion and the longitude of the ascending node
to systematically increase with time in the course of that ``swarm's''
observed duration of about two years.

The experience with the 2010/2011 swarm of bright dwarf sungrazers and comet
C/2011~W3 shows that it is {\it not advisable to employ suddenly increasing
rates of SOHO/STEREO Kreutz sungrazers\/}, especially the brighter ones, {\it
as a token of an imminent\/} (on a time scale of a year or so) {\it arrival
of a spectacular member of the Kreutz system\/}.  Although it would have
worked, because of a coincidence, in the case of C/2011~W3, a prediction of
this kind --- with no further supporting evidence --- is in fact merely a
product of wishful thinking and very risky.  It appears that the process of
cascading fragmentation, whose influence on the evolution of comets in general
and of the Kreutz system in particular proves overwhelming, makes it virtually
impossible to predict the arrival of a spectacular sungrazer on time scales
of a year or so.

In evolutionary terms, the swarm of bright dwarf sungrazers provides
useful information toward our understanding the Kreutz system's morphology,
just as does C/2011~W3.  They both represent {\it warning signals\/} that
the expected new, 21st-century cluster of spectacular members of the Kreutz
system is approaching.  Even though the swarm and C/2011~W3 are products
of different evolutionary paths, the orbital periods --- directly computed
for C/2011~W3 and indirectly inferred for the swarm --- consistently show
that the 20th-century cluster, which was observed to extend from 1945 to
1970, did not represent the last ``bead'' of the ``string of pearls''
[to euphemistically express the extent of the protofragment's perihelion
breakup in analogy to the appearance of comet D/1993~F2 (Shoemaker-Levy)],
but that one or more ``beads'' are yet to come.  The alert to the expected
new cluster is the only parallel link that we find in our effort to decode
the meaning of the two remarkable, nearly-simultaneous events.

\section{Conclusions}

Our study of the population of SOHO/STEREO's Kreutz sungrazers and their
rate of arrival in the years 2004--2013 offers the following conclusions
on the Kreutz system and on the potential temporal relationship with comet
C/2011~W3:

(1) In addition to its usual annual periodicity, caused by selection
effects, the overall arrival-time distribution displays a number of prominent
spikes, whose magnitude peaked approximately one year before the appearance
of comet C/2011 W3, in December 2010.

(2) Comet C/2011 W3 arrived also about one year after the sharply elevated
arrival rate of the SOHO/STEREO sungrazers brighter at maximum light than
apparent magnitude~3, which are free from major selection effects.  A
polynomial fitted through a set of 19 objects of this swarm showed a
peak rate of 4.6 per year centered on 2010.88, while a similar polynomial
fitted to a subset of 13 sungrazers brighter at maximum light than apparent
magnitude~2 indicated a peak rate of 4.3 per year centered on 2011.35,
compared with a pre-2006 average rate of $\sim$1 per year.  The estimated
uncertainty in the times of peak arrival rate is a few months.

(3) A potential physical correlation between comet C/2011~W3 and the elevated
rate of the distribution of bright SOHO/STEREO sungrazers is strongly
contradicted by dynamical evidence, which shows no orbital similarity
between these bright objects and C/2011 W3, with deviations of 10$^\circ$
or more in each of the three angular elements.

(4) On the other hand, fully 18 of the 19 bright SOHO/STEREO sungrazers were
found to move in orbits very similar to that of the spectacular Kreutz comet
C/1843~D1, and 16 of them made up a swarm of tightly related objects, with
clear systematic trends in the argument of perihelion, the longitude of the
ascending node, and the perihelion distance.

(5) The cumulative distribution of SOHO/STEREO sungrazers' arrivals
exhibits a series of flat segments or ``flats,'' relatively prolonged
intervals of time (3--5 weeks) during which the arrival rate of the
population of Kreutz minicomets was nil or extremely low (gaps); they
do not correlate well with the instances of SOHO's roll.

(6) The temporal distribution of the flats is strongly nonuniform:\
no flat is located from the beginning of 2004 to mid-2006, while their
rate is the highest and equal to 3 per year between mid-2006 and mid-2008;
from mid-2009 to mid-2011 the rate is nearly constant at 2 per year.

(7) The number of flats appears to correlate with the arrival rate of
SOHO/STEREO sungrazers brighter at maximum light than magnitude 3; this
correlation appears counterintuitive in that there are no flats in a segment
of the distribution curve that is not populated by any bright dwarf sungrazer
while the temporal rate of flats peaks in a segment of the curve that is
marked by the arrivals of three such sungrazers.

(8) This peculiar property is understood in the framework of cascading
fragmentation, as gaps in the temporal distribution of subfragments are
byproducts of relatively more sizable and cohesive parent fragments; their
breakup generally leads to more nonuniform, cluster-like temporal
arrangement of the ensuing subfragments.

(9) Overall, the results of this study vividly illustrate the
immense morphological complexity of the Kreutz system.  In connection with
C/2011~W3 they not only show that no bright SOHO/STEREO sungrazer was
moving in a similar orbit at least between 2004 and the end of 2012, but
that even faint Kreutz minicomets in such orbits were very rare, a few
close short-lived companions accompanying the comet notwithstanding.

(10) As a plausible interpretation of the evidence, we propose that another
Lovejoy-sized sungrazer, moving in an orbit similar to that of C/1843~D1
but with the perihelion time close to that of C/2011 W3, continued to break
up at large heliocentric distances; and that the swarm of bright dwarf
sungrazers with their arrival rate peaking some time in late 2010 or early
2011 was the prime product of this fragmentation process.

(11) The prominently spike-like distribution of fainter SOHO/STEREO minicomets
suggests that it contains a debris of prematurely fragmented bright dwarf
sun\-{\nopagebreak}grazers also belonging to the swarm.  This scenario is shown to apply to
the most prominent clump of 22 faint SOHO/STEREO sungrazers from mid-December
2010, up to 19 of which fit a test of their association with the swarm.  This
correlation should also be verified on another clump of faint sungrazers from
December 2011, when their orbital data become available.

(12) A decisive role of sheer chance in the timing of the appearance of
comet C/2011~W3, on the one hand, and of the significantly elevated
arrival rate of the population of SOHO/STEREO Kreutz minicomets and
especially the swarm of bright dwarf sungrazers, on the other hand,
is under the circumstances indisputable.

(13) A common origin of all 16 members of the tightly-knit swarm is understood
in terms of perturbations during their original parent's fragmentation, which
proceeded in a cascading fashion at large heliocentric distances.  This
scenario is strongly corroborated by comparison of systematic trends in
the swarm members' orbital elements with the perturbations expected to be
caused by separation velocities acquired by the fragments during such
breakup events.

(14) Spikes in, and sharply elevated rates of, the temporal distribution of
SOHO/STEREO Kreutz sungrazers, bright and faint alike, should not be mistaken
for a token of an imminent (on a time scale of a year or so) arrival of a
spectacular member of the Kreutz system.  Because of an overwhelming effect
of cascading fragmentation, it is practically impossible to predict the arrival
of a major Kreutz sungrazer on such short time scales.

(15) The swarm of bright dwarf sungrazers and the clumps of fainter
sungrazers, causing prominent spikes in the SOHO/STEREO arrival-time
distribution, represent another warning signal that the expected
21st-century cluster of spectacu\-lar Kreutz comets~is~on its way to
perihelion, to arrive in the coming decades (Sekanina \& Chodas 2012).
This alert is the only common link that we find between the recent
nearly-simultaneous appearance of C/2011~W3 and the elevated arrival
rates of the SOHO/STEREO Kreutz population.\\[-0.1cm]

This research was carried out in part at the Jet~Pro\-pulsion Laboratory,
California Institute of Technology, under contract with the National
Aeronautics and Space Administration.\\[-0.25cm]
\begin{center}
{\footnotesize REFERENCES}
\end{center}
{\footnotesize
\parbox{8.6cm}{Brueckner, G. E., Howard, R. A., Koomen, M. J., et al. 1995,
 Sol. {\hspace*{0.4cm}}Phys., 162, 357}\\[0.06cm]
\parbox{8.6cm}{Gray, W. J. 2013, Minor Planet Circ. 84616--84626}\\[-0.03cm]
\parbox{8.6cm}{Green, D. W. E. 2011, IAU Circ. 9246}\\[0.03cm]
\parbox{8.6cm}{Howard, R. A., Moses, J. D., Vourlidas, A., et al. 2008,
 Space Sci. {\hspace*{0.4cm}}Rev., 136, 67}\\[0.04cm]
\parbox{8.6cm}{Knight, M. M., A'Hearn, M. F., Biesecker, D. A., et al.
 2010, AJ, {\hspace*{0.4cm}}139, 926}\\[0.01cm]
\parbox{8.6cm}{Kreutz, H. 1901, Astron. Abh., 1, 1}\\[-0.03cm]
\parbox{8.6cm}{Marsden, B. G. 1967, AJ, 72, 1170}\\[-0.03cm]
\parbox{8.6cm}{Marsden, B. G. 1989, AJ, 98, 2306}\\[0.05cm]
\parbox{8.6cm}{Marsden, B. G., \& Williams, G. V. 2008, {\it Catalogue of
 Cometary {\hspace*{0.4cm}}Orbits 2008} (17th ed.; Cambridge, MA: Minor Planet
 Center \& {\hspace*{0.4cm}}Central Bureau for Astronomical Telegrams,
 195pp)}\\[0.02cm]
\parbox{8.6cm}{Sekanina, Z. 2002, ApJ, 566, 577}\\[-0.03cm]
\parbox{8.6cm}{Sekanina, Z. 2003, ApJ, 597, 1237}\\[-0.03cm]
\parbox{8.6cm}{Sekanina, Z., \& Chodas, P. W. 2007, ApJ, 663, 657}\\[-0.03cm]
\parbox{8.6cm}{Sekanina, Z., \& Chodas, P. W. 2008, ApJ, 687, 1415}\\[-0.03cm]
\parbox{8.6cm}{Sekanina, Z., \& Chodas, P. W. 2012, ApJ, 757, 127 (33pp)}}
\end{document}